\documentclass[runningheads]{llncs}
\usepackage{amsmath}
\usepackage{amsfonts}
\usepackage{setspace}
\usepackage{url}
\usepackage{graphicx}
\usepackage{float}
\usepackage{pgfplots}
\usepackage{algorithm}
\usepackage{tabulary}
\usepackage{comment}
\usepackage[none]{hyphenat}
\usepackage{algorithmic}
\usepackage{chronosys}
\usepackage{multirow}
\usepackage{xcolor} 
\newcommand{\red}[1]{\textcolor{red}{#1}} 
 \newtheorem{thm}{Theorem}[section]

 \newtheorem{defn}[thm]{Definition}

  \numberwithin{equation}{section}

\def\E{\mathbb E}

\def\I{\mathbb I}
\def\N{\mathbb N}

\def\R{\mathbb R}

\newcommand{\CC}{\mathcal{C}}
\newcommand{\CD}{\mathcal{D}}

\newcommand{\CF}{\mathcal{F}}

\newcommand{\CH}{\mathcal{H}}

\newcommand{\CM}{\mathcal{M}}
\newcommand{\CN}{\mathcal{N}}

\newcommand{\CP}{\mathcal{P}}

\newcommand{\CR}{\mathcal{R}}

\newcommand{\CX}{\mathcal{X}}
\newcommand{\CY}{\mathcal{Y}}

\newcommand{\Lap}{\text{Lap}}

\newcommand{\datdom}{\mathbb{N}^{|\CX|}}

 \newcommand{\eps}{\varepsilon}

\newcommand{\be}{\begin{enumerate}}
\newcommand{\ee}{\end{enumerate}}

\begin{document}
\title{Advances in Differential Privacy and Differentially Private Machine Learning}
%
%
\author{Saswat Das\inst{1}\orcidID{0000-0002-6126-1699} \and
Subhankar Mishra\inst{1}\orcidID{0000-0002-9910-7291}}
\authorrunning{Das and Mishra}
%
\institute{National Institute of Science Education and Research, an OCC of Homi Bhabha National Institute, India
\email{\{saswat.das,smishra\}@niser.ac.in}}
\maketitle              
\begin{abstract}
There has been an explosion of research on differential privacy (DP) and its various applications in recent years, ranging from novel variants and accounting techniques in differential privacy to the thriving field of differentially private machine learning (DPML) to newer implementations in practice, like those by various companies and organisations such as census bureaus. Most recent surveys focus on the applications of differential privacy in particular contexts like data publishing, specific machine learning tasks, analysis of unstructured data, location privacy etc. This work thus seeks to fill the gap for a survey that primarily discusses recent developments in the theory of differential privacy along with newer DP variants, viz. Renyi DP and Concentrated DP, novel mechanisms and techniques, and the theoretical developments in differentially private machine learning in proper detail. In addition, this survey discusses its applications to privacy-preserving machine learning in practice and a few practical implementations of DP.

\end{abstract}
\section{Introduction}

The explosion of popularity and adoption of fields like machine learning and big data, and powerful data processing machinery has meant that high quality data is considered to be among the most valuable, high utility commodities. This data, which often includes sensitive details about certain individuals and entities, helps demographers draw useful information about a population and socioeconomic distribution across an area of land, helps tech companies analyse the usage habits of and issues faced by users to design updates to their products, and helps medical professionals to improve upon diagnostic systems and medical care, to understand diseases better, create medical data visualisations, etc. Companies like Netflix and YouTube often utilise data to provide personalised content recommendations for their users. 

But as a corollary, this has enabled 
the extraction of certain, potentially sensitive, information about the individuals in databases (a.k.a. \emph{data subjects}) unless protected in some form.  
{This sensitive information can be used to the detriment of the concerned data subjects by entities like insurance companies that could use sensitive data on whether someone has a particular ailment or habit to increase their insurance premiums or deny them insurance and thus violate legislations like HIPAA that deal with such concerns about sensitive medical data, by other individuals or agencies to blackmail them or track their activities/movements, by governments or political agencies to gain sensitive data on citizens etc.} This has naturally led to privacy concerns. {Well known attacks like the linkage attack on the medical records released by the Massachusetts Group Insurance Commission\cite{Sweeney2002kAnon}, and that that on the Netflix Prize database\cite{nar_shmat_netflix} respectively compromised the medical records of government employees in the state of Massachusetts in the 1990s and the private content consumption data of Netflix viewers in 2006. Very prominently the reconstruction attack on the 2010 US Census data\cite{GAM18} was able to reconstruct the private microdata of a significant proportion of American citizens from deidentified and publicly available census data. Kasiviswanathan, Rudelson, and Smith\cite{kasiviswanathan2012power} (2012) demonstrated that linear reconstruction attacks can be successful in various, including some seemingly \lq\lq non-linear", settings, including when applied to a large class of $ERM$ algorithm outputs, including linear regression and logistic regression. In addition to these well known attacks, more recent ones like that on Myki transportation records released by Public Transport Victoria in 2018\cite{culnane_rubinstein_teague_2019} alarmingly showed that it was possible to trace an individual movements on Melbourne's transportation from these deidentified records by just using two randomly selected touch events/data points.}

  This naturally leads to questions {such as what is} individual privacy, how it can be breached, and how one can go about protecting it while still allowing analysts to draw useful conclusions. In the 1970s, statistical agencies and computer scientists proposed a set of privacy-preserving techniques collectively known as \emph{statistical disclosure limitation} (SDL); this called for \lq\lq anonymising data via methods like top-coding, noise injection into certain attributes in the database, and swappng of attributes among rows. More recent techniques have been proposed, either in light of incidents like these, or in anticipation of certain privacy threats. For instance, the linkage attack on the Massachusetts Group Insurance Commission's medical records led to $k$-anonymity \cite{samarati_k_anonymity}\cite{Samarati98protectingprivacy} being introduced by Samarati and Sweeney to allow for a level of indistinguishability within a database. $k$-anonymity however happens to be vulnerable to certain linkage attacks anyway; and therefore sophisticated variants of $k$-anonymity were introduced subsequently to improve on it, viz. $t$-closeness, $m$-invariance, or $l$-diversity.  {Even so, these are still vulnerable to composition attacks\cite{ganta2008composition}.}

  {However, no notion of privacy provided for an objective assessment of the privacy guarantee granted in the process, and any guarantee thus granted by previously existing techniques or definitions of privacy differed among various contexts. In addition, most of the aforementioned techniques suffered from certain vulnerabilities.}

This state of affairs continued until a powerful class of attacks, known as reconstruction attacks, was introduced by Dinur and Nissim in 2003\cite{dinur2003reconstruction}. Studying these attacks led directly to the introduction of \emph{differential privacy} (often abbreviated as DP) by Dwork et al in 2006\cite{Dwork_McSherry_Nissim_Smith_2006}, often called the gold standard of privacy 
, which provided for a mathematically precise and quantifiable notion of privacy. This notion rests on the idea given by the Fundamental Theory of Information Recovery\cite{DworkRothDP}, which states that providing overly accurate responses to too many queries will inevitably lead to a privacy catastrophe. Thus, differential privacy entails the protection of individual privacy to a large extent by perturbing responses to queries made on a database while still allowing high accuracy of responses and subsequent analysis. The privacy guarantees themselves can be shown in an objective and mathematically rigorous manner.

\subsection{Related Work}

There is a rich trove of literature on differential privacy, various facets of which have been well studied and surveyed in the past. Table \ref{tab:comparison} compares different surveys with respect to ours.

There have been surveys and detailed tutorials that address differential privacy from a technical point of view, like ones by Dwork\cite{Survey:DworkDP}, Dwork and Roth\cite{DworkRothDP}, Vadhan\cite{Vadhan2017Complexity} and \cite{kamath2020primer} which focus on a rigorous treatment of differential privacy; in particular tutorials/textbooks like \cite{DworkRothDP} and \cite{Vadhan2017Complexity} focus mostly on providing a rigorous introduction to the principles of differential privacy, and the basic results, ideas, mechanisms pertaining to it. 

Sengupta et al\cite{NISERDPRev} presented a more recent survey in 2020 which discusses the theory and ideas behind differential privacy, learning with differential privacy and a few industrial deployments of the same. Wang et al\cite{Survey:LDPWang} present an in depth and specific survey on local differential privacy, the theory behind it, its mechanisms and models, variants etc. Xiong et al \cite{Survey:Xiong2020} published a survey on local differential privacy in 2020, taking an general and fairly technical look at local differential privacy, pertinent mechanisms, and its applications to privacy-preserving statistical analysis, machine learning, practical deployments, etc. Fioretto et al\cite{Survey:Fioretto2022DPFairness} (2022) discuss the interplay between differential privacy and fairness, including in the context of differentially private machine learning or decision-making using differential privacy, and how differential privacy affects fairness in these settings. Boulemtafes et al\cite{Boulemtafes2020ARO} (2020) presented a survey on how to train and release deep learning models in a privacy preserving manner in general, and without specific focus on differential privacy.

Other related surveys address different facets of privacy, or that in certain contexts. Dwork et al in \cite{ExposedDwork} discuss attacks on databases, chiefly reconstruction attacks and tracing attacks, and further motivate the need for privacy-preserving data analysis. Sarwate and Chaudhuri's survey\cite{Survey:sarwate2013signal} focuses on differentially private techniques for continuous data for use in signal processing and machine learning. Cunha et al\cite{CUNHA2021100403} present a survey of privacy preserving mechanisms in general, and  Liu et al\cite{Survey:MLmeetsPrivacy} discuss privacy in machine learning in general, but only these surveys only discuss differentially private methods briefly. Gong et al \cite{Survey:GongDPML}(2020) presented a high-level discussing differentially private machine learning exclusively, without delving into technical details rigorously. Zhang et al\cite{Survey:GameTheoryDP} focus on discussing differential privacy in conjunction with game theory, emphasising on game-theoretic solutions to various problems and game-theoretic mechanism design.   {Zhao and Chen (2021)\cite{Survey:ZhaoDPUnstructured} discuss various applications of differential privacy for privacy preserving analysis of unstructured data, viz. images, video, audio etc.}

Surveys by Jiang et al\cite{Survey:LocationPrivacy} and Fung et al\cite{Survey:PrivateDataPublish} discuss location privacy and privacy-preserving data publishing respectively, and include brief discussions on differential privacy as a tool in said contexts.


\begin{table}[t]
\tiny
\centering
\begin{tabulary}{0.8\textwidth}{|p{0.2\linewidth}|>{\centering\arraybackslash}p{0.075\linewidth}|>{\centering\arraybackslash}p{0.075\linewidth}| >{\centering\arraybackslash}p{0.075\linewidth}|>{\centering\arraybackslash}p{0.075\linewidth}|p{0.3\linewidth}|}
\hline
\multicolumn{1}{|c|}{\textbf{Survey}} & \textbf{NoVR} & \textbf{DPML}             & \textbf{NoDP}  & \textbf{CoUs} & \textbf{Remarks/Focus} \\ \hline
Dwork (2008)                                 &               $\times$               &    $\times$                                                 &     $\times$                      &           $\times$     &      First survey on early advances in DP  \\
\hline
Fung et al (201)                                 &               $\times$               &    $\times$                                                &     $\times$                      &           \checkmark     &     Privacy in Data-Publishing  \\
\hline
Sarwate and Chaudhuri (2013)                                 &               $\times$               &    \checkmark                                                &     $\times$                      &           \checkmark     &     DP in Signal Processing and ML  \\
\hline
Dwork and Roth (2014)                                 &               $\times$               &    \checkmark                                                 &     $\times$                      &           $\times$             & Seminal primer on DP discussing various aspects of it and relevant algorithms and applications \\ \hline
Vadhan (2016)                                 &               $\times$               &    $\times$                                                 &     $\checkmark$                      &           $\times$         &   Theoretically rigorous \\ \hline
Dwork et al (2017)                                 &               $\times$               &    $\times$                                                 &     $\times$                      &           \checkmark         &   Privacy attacks on databases \\ \hline
Wang et al (2020)                                &           \checkmark                     &              \checkmark             &        \checkmark                 &      \checkmark       & LDP   \\ \hline
Xiong et al (2020)                                &           \checkmark                     &              \checkmark             &        \checkmark                 &      \checkmark       & LDP   \\ \hline
Sengupta et al (2020)                                &           $\times$                     &              \checkmark             &        \checkmark                   &      $\times$     &   Learning with DP, Implementations  \\ \hline
Gong et al (2020)                                &           $\times$                     &              \checkmark             &        $\times$                   &      $\checkmark$     &  High-level discussion on applied DPML   \\ \hline
Boulemfates et al (2020)                                &           $\times$                     &              $\checkmark$             &        $\times$                   &      $\times$     &   General survey on privacy preserving deep learning\\ \hline
Jiang et al (2021)                                &           $\times$                     &              $\times$             &        $\times$                   &      $\checkmark$     &   Privacy Preserving Mechanisms for Location Based Services\\ \hline
Liu et al (2021)                                &           $\times$                     &              $\checkmark$             &        $\times$                   &      $\checkmark$     &   Privacy Preserving ML (not restricted to DPML)\\ \hline
Cunha et al (2021)                                &           $\times$                     &              $\times$             &        $\times$                   &      $\times$     &   Privacy Preserving Mechanisms (including DP) in general\\ \hline
Zhang et al (2021)                                &           $\times$                     &              $\times$             &        $\times$                   &      $\checkmark$     &   Game Theory with DP\\ \hline
Zhao and Chen (2021)                                &           $\times$                     &              $\times$             &        $\times$                   &      $\checkmark$     &   Analysis of Unstructured Data with DP\\ \hline
Fioretto et al (2022)                       & $\times$                             & \checkmark & $\times$ &    \checkmark       &              Intersection of Fairness and DP                                    \\ \hline
\textbf{Das and Mishra}                        & \checkmark                           & \checkmark & \checkmark &    $\times$       & \textbf{Our Survey}                                                 \\ \hline
\end{tabulary}
\caption{Table comparing similar surveys with this work. The adjective \lq\lq novel" in the table refers to anything that has been introduced in or after 2015. A \checkmark is awarded if a survey makes more than a brief mention of a topic and discusses it in some detail. NoVR - Novel Variants of Differential Privacy, DPML - Differential Privacy in Machine Learning, NoDP - Novel DP mechanisms and techniques, CoUs - Context or Usage specific survey} 
\label{tab:comparison}
\end{table}
\subsection{Our Contribution}


The works (viz. \cite{DworkRothDP}) cited above include some seminal ones and remain a vital introduction to the study of differential privacy. But the recent surveys on differential privacy literature that the authors have encountered are either introductory in nature or focus only on a specific context of use or facet of differential privacy, or they treat differential privacy as a supplement for certain application, without paying much attention to the technical aspects of it. Surveys discussing the most novel methods, variants and developments as of the moment do not exist. In addition, newer privacy loss accounting techniques, variants of differential privacy, and several applications to fields like machine learning have been introduced in recent years. This marked a need for a newer, more general, and broader survey on some facets of the rapidly expanding and recent literature on differential privacy,  {in particular novel variants of DP, applications to machine learning and data analysis, tighter bounds and deployments}, while bringing back some emphasis to central differential privacy. This is a need this survey seeks to fulfill.

  

Contributions of our survey: 
\begin{itemize}
    \item We review basic definitions and ideas central to the theory and study of differential privacy briefly for readers who might not be very acquainted with the subject matter.
    \item We then discuss basic mechanisms that are used to implement differential privacy, recent results and privacy analysis, and some variants of differential privacy, the motivation behind them and their salient features.
    \item We discuss the theoretical foundations of differentially private machine learning and deep learning, and novel advances and algorithms in these fields, including in contexts like federated learning.
    \item We also take a look at some industrial/practical deployments of differential privacy to provide an idea of how some of the largest data-intensive companies/agencies use differentially private techniques to preserve privacy of their users' data.
    \item {We perform bibliometric analysis of the papers that have been published in order to give an idea of the directions research in differential privacy is moving in.}
\end{itemize}

\begin{figure}[!h]
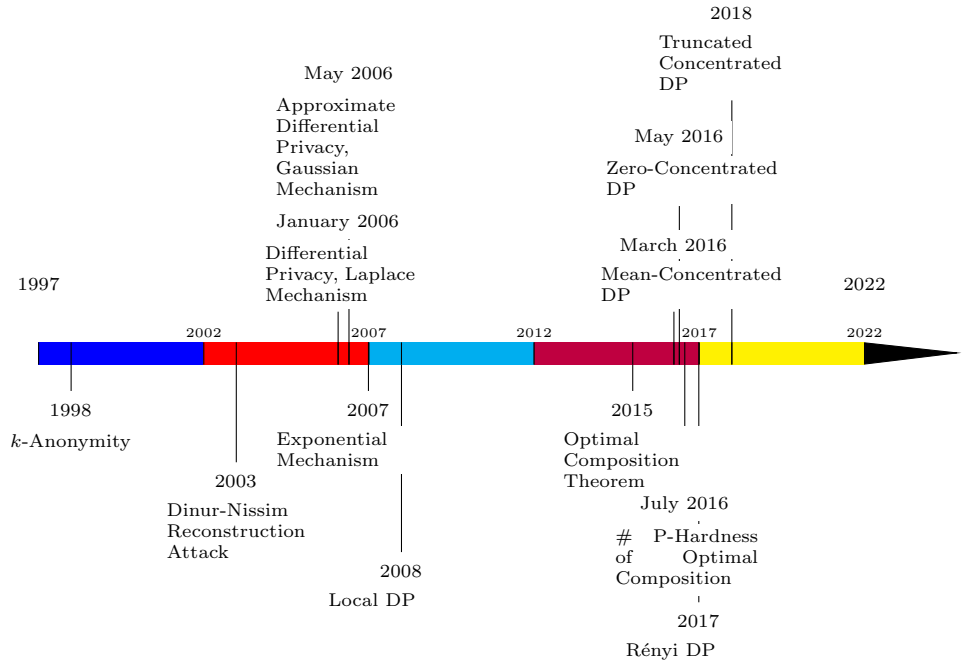

\scriptsize
\catcode`\@=11
\def\chron@selectmonth#1{\ifcase#1\or January\or February\or
March\or April\or May\or June\or July\or August\or
September\or October\or November\or December\fi}
\startchronology[startyear=1997,stopyear=2022]
\chronograduation[periode][dateselevation=0pt]{5}
\chronoevent[textwidth=2cm,markdepth=-105pt]{2018}{Truncated Concentrated DP}
\chronoevent[textwidth=2cm,markdepth=-65pt]{05/2016}{Zero-Concentrated DP}
\chronoevent[textwidth=2cm,markdepth=-25pt]{03/2016}{Mean-Concentrated DP}
\chronoevent[textwidth=2cm,markdepth=90pt]{2017}{R\'enyi DP}
\chronoevent[textwidth=1.9cm,markdepth=45pt]{07/2016}{\# P-Hardness of Optimal Composition}
\chronoevent[textwidth=1.9cm,markdepth]{2015}{Optimal Composition Theorem}
\chronoevent[textwidth=2cm,markdepth=71pt]{2008}{Local DP}
\chronoevent[textwidth=2cm,markdepth=-65pt]{05/2006}{Approximate Differential Privacy, Gaussian Mechanism}
\chronoevent[textwidth=2cm,markdepth=-25pt]{01/2006}{Differential Privacy, Laplace Mechanism}
\chronoevent[textwidth=1.9cm,markdepth=37pt]{2003}{Dinur-Nissim Reconstruction Attack}
\chronoevent[textwidth=2.5cm]{2007}{Exponential Mechanism}
\chronoevent{1998}{$k$-Anonymity}
\chronoevent[textwidth=2.5cm]{1965}{Warner's Randomised Response}
\stopchronology
\caption{Timeline of Important Definitions and Developments }
\label{fig:timeline1}
\end{figure}

In addition, readers are encouraged to refer to Figure \ref{fig:timeline1} for a concise timeline on some important developments in the study of differential privacy.

\section{Definitions, Mechanisms, and Variants}

To concretely discuss the subject matter of this paper, it is imperative to provide a quick and brief introduction to differential privacy and relevant definitions\cite{DworkRothDP}. In particular, this section shall, in a brief technical manner, introduce differential privacy, its salient properties, and some basic tools used to implement. In addition, novel variants of differential privacy, the ideas behind them, and the properties associated with them shall be discussed.



\subsection{Pure and Approximate Differential Privacy and Prerequisite Definitions}
To start off, we shall define what differential privacy is concretely.

$$||x||_p=(\sum_{i=1}^{|\CX|}|x_1|^p)^\frac 1p.$$



\begin{defn}{\sc{$\eps$-Differential Privacy}}\\
A randomised algorithm/mechanism $\CM$ is said to be \emph{$\varepsilon$-differentially private} or \emph{purely differentially private} if $\forall\,S\subseteq\text{Range}(\CM)$ and $\forall\,x,y\in\N^{|\CX|}$ such that $||x-y||_1\leq 1$ (i.e. for \emph{neighbouring databases} or $x\sim y$),
$$\ln\left(\frac{\Pr[\CM(x)\in S]}{\Pr[\CM(y)\in S]}\right)\leq\varepsilon$$
with the probability space being over the coin flips of the mechanism $\CM$.
\end{defn}

The above definition of differential privacy is also the earliest, and was given by Dwork, McSherry et al\cite{Dwork_McSherry_Nissim_Smith_2006} in 2006. A relaxation of this was given by Dwork et al\cite{OurDataDwork} shortly later in 2006 in order to provide comparable, albeit slightly weaker privacy guarantees, with addition of significantly less noise, and excusing events that have low probability (denoted by $\delta$) of occurring.

\begin{defn}{\sc{$(\eps,\delta)$-Differential Privacy}}\\
A randomised algorithm $\CM$ on the domain $\N^{|\CX|}$ is said to be $(\varepsilon,\delta)$-differentially private (or \emph{approximately differentially private} if $\delta>0$) if $\forall\,S\subseteq\text{Range}(\CM)$ and $\forall\,x,y\in\N^{|\CX|}$ such that $||x-y||_1\leq 1$ (i.e. for \emph{neighbouring databases}),
$$\ln\left(\frac{\Pr[\CM(x)\in S]-\delta}{\Pr[\CM(y)\in S]}\right)\leq\varepsilon$$
with the probability space being over the coin flips of the mechanism $\CM$.
\end{defn}


\begin{figure}
    \centering
    \includegraphics[width=\textwidth]{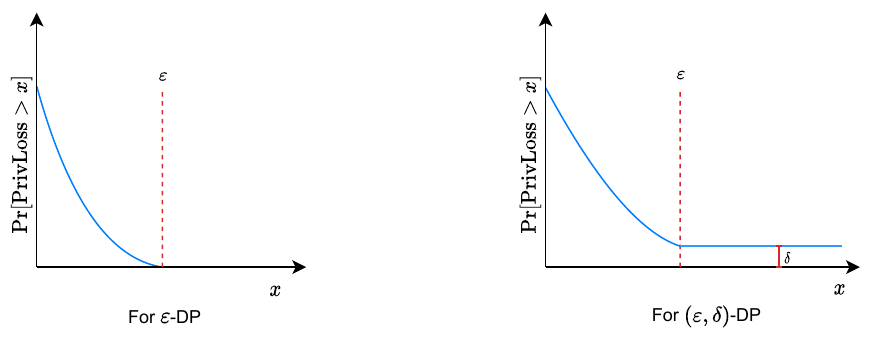}
    \caption{Privacy Loss Graphs for Pure and Approximate Differential Privacy.}
    \label{fig:privloss1}
\end{figure}
The privacy loss graphs for $\eps$-DP and $(\eps,\delta)$-DP are given in Figure-\ref{fig:privloss1}.
\subsubsection{Basic Properties of $(\eps,\delta)$-DP}
Differential privacy possesses some very appealing properties that make modular design of differentially private algorithms and the analysis of their privacy properties possible, all of which are mentioned and proven in \cite{DworkRothDP}. These properties are as follows.

\begin{thm}{\sc{Post-Processing Invariance}}\\
Let $\CM:\N^{|\CX|}\to R$ be a randomised algorithm that is $(\varepsilon,\delta)-$differentially private. Let $f:R\to R'$ be an arbitrary randomised mapping. Then $f\circ\CM:\N^{|\CX|}\to R'$ is $(\epsilon,\delta)-$differentially private.
\end{thm}

In simpler terms, post-processing invariance simply means that an adversary cannot degrade the privacy of the output of a differentially private algorithm by post-processing it by any means without using the raw data itself\cite{Survey:sarwate2013signal}.

For an algorithm satisfying $\eps$-DP for databases that are at most $k$ distance apart, \emph{group privacy} provides a bound of $k\eps$ on the privacy loss.
\begin{thm}{\sc{Group Privacy}}\\
Let $\CM:\N^{|\CX|}\to R$ ($R$ being an appropriate codomain/range of the mechanism) be an $(\varepsilon,\delta)-$differentially private mechanism; for groups of size $k$, i.e. $\forall$ databases $x,y$ such that $||x-y||_1\leq k$ and all $T\subseteq R$,
$$\ln\left(\frac{\Pr[\CM(x)\subseteq T]-\delta'}{\Pr[\CM(y)\subseteq T]}\right)\leq k\varepsilon.$$
Where $\delta'=\frac{e^{k\cdot\eps}-1}{e^\eps-1}\cdot\delta$.

When $\delta=0$, the mechanism satisfies $\eps$-DP, and for groups of size $k$, the above reduces to
$$\ln\left(\frac{\Pr[\CM(x)\subseteq T]}{\Pr[\CM(y)\subseteq T]}\right)\leq k\varepsilon.$$
\end{thm}

An often studied aspect of DP involves the composition of DP algorithms. This is based on the fact that DP allows graceful composition of various $(\eps_i,\delta_i)$-DP algorithms to provide $(\sum_i\eps_i,\sum_i\delta_i)$-DP.
\begin{thm}{\sc{Composition of Differentially Private Mechanisms}}\\
Let $\{\CM_i\}_{i\in[k]},$ where $\CM_i:\datdom\to\CR_i$, be a finite sequence of \\$(\varepsilon_i,\delta_i)-$differentially private algorithms. If $$\CM_{[k]}:\datdom\to\prod_{i\in[k]}\CR_i, \CM_{[k]}(x)=(\CM_1(x),\ldots,\CM_k(x)),$$ then $\CM_{[k]}$ is $\left(\sum_{i=1}^k\varepsilon_i,\sum_{i=1}^k\delta_i\right)-$differentially private.  
\end{thm}
\subsubsection{Tighter Composition Bounds}
There have been attempts to further refine the composition bound. The advanced composition theorem, due to Dwork, Rothblum, and Vadhan\cite{AdvCompBoosting}, provides a tighter bound on composition under $k$-fold adaptive composition.

\begin{defn}{\sc{$k$-Fold Adaptive Composition}\\}
For $b\in \{0,1\}$, family $\CF$, and adversary $A$, for each $i\in[k]$, $A$ produces two neighbouring databases $x_i^0$ and $x_i^1$, a mechanism $\CM_i\in\CF$, and parameters $w_i$, and is returned a randomly chosen $y\in\CM_i(w_i,x_i^b)$.\\
The choice of $b$, once made is kept constant throughout the experiment (ergo giving us two different variants of the experiment).
\end{defn}

Under $k$-fold adaptive composition, the advanced composition theorem is stated as follows.\footnote{We shall forego a detailed discussion on $k$-fold adaptive composition and the advanced composition theorem for the sake of brevity. Details can be found in the cited paper and \cite{DworkRothDP}.}

\begin{thm}{\sc{Advanced Composition for Differentially Private Mechanisms}\\}
$\forall\,\eps>0,\delta,\delta'\in[0,1]$, and $k\in\N$, the class of $(\eps,\delta)$-differentially private mechanisms is \mbox{$(\eps',k\delta+\delta')$} differentially private under $k$-fold adaptive composition, where
\[\eps'=\sqrt{2k\ln(1/\delta')}\cdot\eps+k\eps(e^\eps-1).\]
\end{thm}

In 2015, Kairouz, Oh and Viswanath\cite{kairouz2015optimal} gave the optimal composition theorem, which provides a tighter composition bound than the advanced composition theorem itself under $k$-fold adaptive composition. The statement of the theorem in \cite{kairouz2015optimal} is quite verbose. A simpler, but equivalent, restatement of the same was given by Murtagh and Vadhan\cite{murtagh2016complexity}, which is given below.

\begin{thm}{\sc{Optimal Composition Theorem}\\}
For all $\eps_i>0$ and $\delta_i\in[0,1)$, where $i\in[k]$, and for any $\delta'\in[0,1)$, the composition of the algorithms/mechanisms $\CM_1,\CM_2,\ldots,\CM_k$, where $\CM_i$ is $(\eps_i,\delta_i)$ differentially private, yields $(\eps',\delta')$-differential privacy with the least value of $\eps'>0$ satisfying
\[\frac1{\prod_{i\in[k]}(1+e^{\eps_i})}\sum_{S\subseteq[k]}\max\left\{e^{\sum_{i\in S}\eps_i}-e^{\eps'}\cdot e^{\sum_{i\notin S}\eps_i},0\right\}+\frac{1-\delta'}{\prod_{i\in[k]}(1-\delta_i)}\leq 1.\]
\end{thm}

Murtagh and Vadhan\cite{murtagh2016complexity} showed that computing optimal composition is \# P hard, even under simpler conditions like when only purely differentially private mechanisms are being composed. This makes it a computationally difficult problem. Also note that the composition of approximate differentially private algorithms seems to pay a penalty in factors of $\log(\frac1\delta)$; each step of composition entails a $\sqrt{\log\left(\frac 1\delta\right)}$ penalty on privacy loss.

A corollary of the above penalty meant that composition of $(\eps,\delta)$-differentially private algorithms is not associative, by which we mean that the composition of $(\eps,\delta)$-DP algorithms is not independent of the order and manner in which it is done. 
\vspace*{-10pt}
{\begin{table}[!h]
\footnotesize 
\begin{tabular}{|p{0.2\linewidth}|p{0.2\linewidth}|p{0.2\linewidth}|p{0.3\linewidth} | p{0.25\linewidth}|}
\hline
\multicolumn{1}{|c|}{\textbf{Variant}} & \textbf{Notation}  & \textbf{Canonical Mechanism} & \textbf{Advantages or Improvements}                                       & \textbf{Disadvantages}                                                                                        \\ \hline
Pure DP                                         (Jan, 2006)              & $\eps$-DP          & Laplace            & Elegant Composition, Group Privacy, Post Processing Invariance               & Too strict, requires significant noise addition at times                                                      \\ \hline
Approximate DP                                  (May, 2006)              & $(\eps,\delta)$-DP & Gaussian           & Relaxes pure DP, requires less noise addition                             & Can lead to catastrophic failure with a small probability $\delta$; Composition is not elegant or associative \\ \hline
Mean Concentrated DP \cite{dwork2016concentrated}                           (2016)            & $(\mu,\tau)$-mCDP  &     Gaussian                         & Avoids catastrophic failure, tighter composition bounds, concentrates privacy loss around a bounded mean, associative and elegant composition & Not post-processing invariant                                                                                 \\ \hline
Zero Concentrated DP    \cite{bun2016concentrated}                        (2016)            & $(\eta,\rho)$-zCDP  & Gaussian                             & Preserves the benefits of mCDP, is post-processing invariant, avoids catastrophic failure, associative and elegant composition &          Tools like Propose-Test-Release and Amplification via Subsampling not supported.                                                                        \\ \hline
Truncated Concentrated DP              \cite{bun2018truncated}              (2018)            & $(\rho,\omega)$-tCDP  & $\sinh$-Normal                             & Relaxes zCDP, supports techniques like Propose-Test-Release and Amplification via Subsampling &                                                                          N/A        \\ \hline
R\'enyi DP        \cite{MironovRenyiDP}                    (2017)            & $(\alpha,\eps)$-RDP  & Gaussian                              & No catastrophic failure, tighter error bounds, better privacy loss accounting for approximate DP, linearly additive composition &                                                                            Tighter privacy bound on Gaussian is possible      \\ \hline
Gaussian DP            \cite{Dong2020GaussianDP}                (2020)            & $\mu$-GDP  & Gaussian                              & Tightest possible privacy bound for the Gaussian mechanism, tight composition &                                                                            N/A      \\ \hline
\end{tabular}
\caption{Table summarising different variants of (central) differential privacy \red{
}
}
\label{tab:DPVariants}
\end{table}}

These are certain issues that demanded attention, and were ultimately dealt with with the introduction and study of certain, more novel variants of differential privacy. {Owing to size constraints, those are very briefly discussed here and in table \ref{tab:DPVariants}.}

{\emph{Concentrated differential privacy} (CDP) was introduced and discussed by Dwork and Rothblum\cite{dwork2016concentrated} (mCDP), and further refined by Bun and Steinke\cite{bun2016concentrated} (zero CDP or zCDP) in 2016 in order to address some of the aforementioned issues with existing definitions of DP and to gain sharper bounds on and associativity of composition. These are defined in terms of R\'enyi divergence\cite{renyi1961entropy} and the privacy loss random variable $Z$, and demand that $Z$ is \emph{concentrated} around a bound on the mean of the privacy loss $\mu$ and zero respectively. Interestingly, the optimal composition of zCDP can be computed in linear time, in contrast to that of $(\eps,\delta)$-DP, which is \#P hard.}

{However, some tools like propose-test-release and amplification-via-subsampling are not supported by zCDP, which led Bun, Dwork et al\cite{bun2018truncated} to introduce a relaxation on zCDP, called truncated CDP in 2018. while CDP demands that the privacy loss is at least as concentrated as a Gaussian, tCDP relaxes that demand to having the privacy loss being concentrated like a Gaussian up until a certain amount of standard deviations (roughly $\omega$) away.}

{\emph{R\'enyi differential privacy} was introduced by Mironov\cite{mironov2019renyi} in 2017, which is also defined with respect to R\'enyi divergence. This seeks to improve on $(\eps,\delta)$-DP by modifying the relaxation condition from having a potential catastrophe occur with probability $\delta$ to weakening the DP guarantee in another fashion, and to provide better privacy loss accounting for the Gaussian mechanism. Moreover, Geumleuk, Song, and Chaudhuri \cite{GeumlekSongChaudhuri} (2017) stated that many differentially private mechanisms that sample from distributions from exponential families, viz. posterior sampling, have closed-form R\'enyi DP guarantees available.}

{However, the aforementioned variants face issues regarding the composition of private algorithms and the analysis of techniques such as privacy amplification via subsampling\cite{balle2018privacy}. To remedy that, Dong, Roth and Su\cite{Dong2020GaussianDP} (2020) introduce a family of definitions of DP called $f$-differential privacy. This definition is motivated by differential privacy being formulated as a hypothesis testing problem for an adversary by \cite{Wasserman2008ASF} and \cite{kairouz2015optimal}. The authors also defined a specialisation of $f$-DP called Gaussian differential privacy or $GDP$.}
\subsection{Basic Mechanisms}
Bringing differential privacy from the realm of theory to practice involved the introduction of various mechanisms used to achieve it in different ways and contexts. This subsection shall deal with some fundamental DP mechanisms. Most mechanisms endow DP guarantees on data releases by perturbing the data in some form or fashion, including by adding noise from an appropriate distribution. This perturbation can be done by perturbing the raw data prior to answering a query made on the dataset, or by perturbing the query response received on the raw data.

The earliest model of applying differential privacy involves having a trusted curator who holds all the data of the data subjects and is responsible for responding to queries made by analysts in a way that upholds differential privacy of the data. This is called \emph{central differential privacy} (CDP). CDP is often enforced by the curator by the addition of noise to the data to perturb the values of true query responses by the use of privacy mechanisms.

There are some well known mechanisms used to add noise to query responses via different methods, the oldest of which is the global sensitivity method, which calibrates the noise added to responses to numeric queries with respect to a quantity called the \emph{global sensitivity} of the query. 
\begin{defn}{\sc{Global Sensitivity of (a Set of) Queries}}\\
The \emph{global sensitivity of a query $f$}, or the $\ell_p$ sensitivity of $f$ is given by $$\Delta_p(f)=\max_{\Vert x-y\Vert_p\leq 1}|f(x)-f(y)|,$$ 
and that of a set of queries $Q$ is $$\Delta_1(Q)=\sup_{\Vert x-y\Vert_p\leq 1}\left(\sum_{q\in Q}|q(x)-q(y)|^p\right)^{\frac 1p}.$$
\end{defn}

The Laplace mechanism from a paper by Dwork et al\cite{Dwork_McSherry_Nissim_Smith_2006} (which also provided the first known definition of what we now call pure differential privacy) calibrates the added noise with respect to the $\ell_1$ sensitivity of the query being made, or the set of queries being made.

\begin{defn}{\sc{Laplace Mechanism}}\\
The Laplace distribution, $\Lap(\mu,b)$, is given by the pdf $p(z|\mu,b)=\frac 1{2b}\exp\left(-\frac{|z-\mu|}{b}\right)$.\\
We denote $\Lap(b):=\Lap(0,b)$, pdf, $p(z|b):=p(z|\mu=0,b)=\frac 1{2b}\exp\left(\frac{-|z|}b\right)$.\\
Given $\eps>0$, a set of queries, $Q$, and an input database $x$, the Laplace mechanism $\CM_L$ returns noisy answers $\{q(x)+\Lap(\Delta_1(Q)/\eps)\}_{q\in Q}$.
\end{defn}

Dwork et al\cite{Dwork_McSherry_Nissim_Smith_2006} showed that the Laplace mechanism is $\eps$-differentially private.

Later, the notion of approximate (i.e. $(\eps,\delta)$-) differential privacy was introduced in \cite{OurDataDwork} along with the addition of Gaussian noise as a means of achieving it. The Gaussian mechanism, which is an $(\eps,\delta)$-differentially private mechanism, utilises the global sensitivity method, and adds Gaussian noise calibrated to the $\ell_2$ sensitivity of queries made.

\begin{defn}{\sc{Gaussian Mechanism}}\\
Let $N(\mu,\sigma^2)$ denote the Gaussian distribution with mean $\mu$ and standard deviation $\sigma$.\\
Given some $\eps>0$ and $\delta>0$, a set of queries $Q$, and an input database the Gaussian mechanism $\CM_G$ returns noisy answers $\{q(x)+ N(0,\sigma^2)\}_{q\in Q},$ where $\sigma\geq c\frac{\Delta_2(Q)}{\varepsilon},$ and $c^2>2\ln\left(\frac{1.25}{\delta}\right).$

It is common to take $\sigma^2=2\ln(\frac{1.25}\delta)\frac{\Delta_2(Q)^2}{\varepsilon^2}$.
\end{defn}

Additive noise addition is not suited for some contexts, and might even be counterproductive, like in some contexts involving choosing a best object from a database. For that reason McSherry and Talwar\cite{exp_mech} introduced the exponential mechanism in 2007, which is $\eps$-differentially private.

\begin{defn}{\sc{Exponential Mechanism}}\\
Given a database $x\in\datdom$, a set of objects $\CH$, a score function $s:\datdom\times\CH\to\R$, and $\eps>0$, the exponential mechanism $\CM_E$ outputs $h\in\CH$ with probability $=c \exp(\frac{\varepsilon s(x,h)}{2\Delta s})$, where $c\in\R_+$ is a suitable constant.
\end{defn}
However, in certain contexts, having a trusted curator is not possible or is not desirable, and the data subjects could instead perturb their data locally and respond to queries. This brings us to the following definition given by Kasiviswanathan et al\cite{WhatCanWeLearnKasi}.

\begin{defn}{\sc{Local Differential Privacy (LDP)}\\}
A randomised mechanism $\CM$ for $\varepsilon>0$ is said to be $\varepsilon-$locally differentially private if for all pairs $x,y$ of a user's private data, and for all possible outputs $z\in\text{Range}(\CM)$,
$$\ln\left(\frac{\Pr[M(x)=z]}{\Pr[M(y)=z]}\right)\leq\varepsilon$$
with the probability space being over the coin flips of $\CM$.
\end{defn}

The very first implementation of local differential privacy interestingly vastly predates the conception of differential privacy itself; in 1965, Warner\cite{WarnerRandomResp} came up with the concept of \emph{randomised response} for data collection about sensitive topics. A simple version, as mentioned in \cite{DworkRothDP}, of this involves the use of a fair coin; given a sensitive property $\CP$, a respondent flips a coin. If they obtain a tails, then they answer truthfully, else they flip the coin again; in the latter event, if they obtain a heads, they answer yes, else they answer no. The coin tosses and the number of them remain private to the respondent. This is shown to give a close approximation of the expected number of people who possess the property $\CP$, given by $$\E[\text{Yes}]=\frac 34 n(\text{has }\CP)+\frac 14 n(\text{does not have }\CP).$$

In the context of differential privacy, it is defined for some $\eps>0$ as follows (and thus is $\eps$-locally differentially private).

\begin{defn}{\sc{$\eps$-Locally Differentially Private Randomised Response}}\\
Given $\eps>0$, for every private bit $X$ in a piece of data, output
$$\CM(X)=\begin{cases}X, \text{ with probability}=\frac{\exp{(\varepsilon)}}{1+\exp{(\varepsilon)}};\\
    1-X, \text{ with probability}=\frac{1}{1+\exp(\varepsilon)}.\end{cases}$$
\end{defn}


\section{Differentially Private Algorithms}

While various definitions of differential privacy have been discussed along with some basic mechanisms, differential privacy can be implemented in ways that try to reduce or minimise error and privacy loss, while answering queries optimally, and sometimes only when they satisfy certain conditions. We discuss some prominent techniques, and some research done on and improvements made to them post their initial introduction. 


\subsection{Sparse Vector Technique (SVT)}
The Sparse Vector Technique, or SVT, was introduced by Dwork et al\cite{Dwork2009Complexity} in 2009 and later improved upon by Roth and Roughgarden\cite{roth2011interactive} in 2009 and Hardt and Rothblum\cite{Hardt2010MultWt} in 2010. It serves the purpose of answering only a certain number ($c$) of queries from a sequence of $k$ low sensitivity, and adaptively chosen queries with noise addition given that they lie above a certain threshold, $T$. Therefore the privacy loss would not increase as a factor of $k$ (for pure DP) or $\sqrt k$ (as for approximate DP), but will depend on $c$, where $c\ll k$.  {In fact, the noise added by SVT scales as $\theta(\log k)$.}

  {For an adaptively chosen sequence of $\frac 1n$ sensitivity queries, the original SVT algorithm has $\delta=0$, and noise from $\Lap(\sigma)$ is added to the threshold $T$ to obtain a noised threshold $\hat T$. Laplace noise from $\Lap(2\sigma)$ is added to each query response and compared against the noised threshold, $\hat T$; only if the noised query response exceeds $\hat T$ does the algorithm output $\top$, else it outputs $\bot$, thus identifying \lq\lq meaningful" queries the responses to which exceed the noised threshold. This satisfies $\eps$-DP. Modified SVT follows the same paradigm as above but with $\sigma=\frac{\sqrt{32c\ln\frac 1\delta}}{n\eps}$, and this satisfies $(\eps,\delta)$-DP.}


After obtaining the meaningful queries from SVT, one can get these queries answered using differentially private noise addition mechanisms. Another modification, called NumericSparse, augments Sparse to enable it to release query responses with Laplace noise addition for meaningful queries, and output $\bot$ for all others, with $(\eps,\delta)$ differential privacy. 

However, SVT as initially proposed remains practically infeasible. For instance in 2018, Papernot et al\cite{papernot2018knowledgetransfer} showed that SVT is often outperformed by the Gaussian mechanism answering all queries in model agnostic learning, owing to more concentrated noise addition and tighter composition via CDP or RDP based privacy accounting.

In 2020, Kaplan, Mansour, and Stemmer\cite{kaplan2020svt} devised an improvement to SVT by doing away with the SVT algorithm's reinitialisation after every meaningful query, and by deleting records from the dataset after they have contributed to some $k$ many $\top$ answers. It is noted while SVT can answer $c$ meaningful queries for a database of size$=\tilde O(\sqrt c\cdot\log m)$, Kaplan et al's technique can do the same for a database of size$=\tilde O(\sqrt{k^*}\cdot\log m)$, where $k^*$ is the maximum number of times any record contributes to the response to a meaningful query, and is thus $\leq c$. This was also demonstrated to be useful for the shifting-heavy hitters problem\footnote{An element $x$ is called a heavy hitter if it is current input for a large number of users.}.

Zhu and Wang\cite{Zhu2020RenyiSVT} (2020) studied a generalised family of SVT that allows for the use of any noise-adding mechanisms, and introduced a variant of SVT that uses Gaussian noise instead of Laplace noise. They found tighter RDP bounds for SVT, improving upon previously known bounds by a constant factor.
  
\subsection{Privacy Amplification via Subsampling}
Privacy Amplification via Subsampling promises that applying $(\eps,\delta)$-differentially private mechanisms to random $\gamma$-subsets of records of a dataset yields a stronger privacy guarantee in the form of $(O(\gamma\eps),\gamma\delta))$-DP\cite{balle2018privacy}. The intuitive reason for this is that by picking a random subset/sample from the dataset, we reduce the probability of a data point that differs between that dataset and its neighbouring datasets from appearing in that sample. 

The benefits of using this principle have been widely recognised, often providing significant improvements in terms of privacy loss wherever applied. A variant of the Gaussian mechanism called the Sampled Gaussian Mechanism (SGM) combines Privacy Amplification via Subsampling and the Gaussian mechanism, using which the privacy cost of a single evaluation diminishes quadratically with respect to the sampling rate.\footnote{The following definition is from \cite{mironov2019renyi}.}
\begin{defn}{\sc{Sampled Gaussian Mechanism (SGM)}}\\
Let $f:P(S)\to\R^d$ be a function mapping subsets of a set $S$ to $d$-dimensional real valued vectors. Then the Sampled Gaussian Mechanism is defined with respect to a sampling rate $q\in(0,1]$ and $\sigma>0$ as
\[SG_{q,\sigma}(S):=f(\{x:x\in S\text{ is sampled with probability }q\})+\CN(0,\sigma^2\I_d),\]
Where each element of $S$ is sampled independently at random with probability $q$ without replacement, and $\CN(0,\sigma^2\I_d)$ is spherical $d$-dimensional Gaussian noise with per-coordinate variance $\sigma^2$.
\end{defn}
Mironov et al\cite{mironov2019renyi}(2019) discuss SGM and provide a numerically stable method to calculate the R\'enyi DP of SGM precisely and provide nearly tight closed form bounds on the RDP of SGM.
\begin{thm}{\sc{Closed Form Bound}}\\
If $q\leq\frac 15$, $\sigma\geq4$, and if along with $\alpha$, these satisfy
\[1<\alpha\leq\frac 12\sigma^2 L-2\ln\sigma; \alpha\leq\frac{\frac12\sigma^2L-2\ln\sigma}{L+\ln(q\alpha)+\frac1{2\sigma^2}}\]
where $L:=\ln\left(1+\frac 1{q(\alpha-1)}\right)$, then SGM applied to a function of $\ell_2$-sensitivity $I$ satisfies $(\alpha,\eps)$-RDP where $$\eps:=\frac{2q^2\alpha}{\sigma^2}.$$
\end{thm}

\subsection{Privacy Amplification via Shuffling}
Privacy Amplification via Subsampling\cite{erlingsson2020amplification} seeks to strengthen the privacy guarantees of a locally differentially private algorithm when viewed through the lens of central differential privacy; by privacy amplification via shuffling, a permutation invariant algorithm satisfying $\eps$-LDP can be shown to satisfy $\left(O\left(\eps\sqrt{\frac{\log(1/\delta)}n}\right),\delta\right)$-DP.

Shuffling in itself is a powerful primitive. Techniques like BUDS by Sengupta et al\cite{Sengupta2020BUDS} primarily rely on shuffling techniques to provide differential privacy guarantees. BUDS in particular makes use of a technique the authors call iterative shuffling, which involves randomly choosing a shuffler from a set of shufflers in each iteration and shuffling a given lot of data using it until all of the data has been shuffled. A recent improvement to BUDS, titled BUDS+\cite{sengupta2022budsplus}, was introduced by the authors in 2022, which improves on the privacy utility tradeoff, longitudinal privacy guarantees, with memory efficiency and improved security guarantees.

\section{Applications in Machine Learning} 
{Differential privacy has found a rich potential for use in various realms of machine learning. Non-private machine learning carries certain risks, given that training an ML model uses a large volume of data, and non-private machine learning does not by default ignore specific facts provided in the training data, which can be used to compromise the privacy of the raw training data used to train a given model.}

For example, in 2017, Shokri et al\cite{shokri2017membership} demonstrated a membership inference attack that, given a model, could infer whether an individual was included in the model's training dataset. Fredrikson et al\cite{Fredrikson2014ModelInversion} (2014) and Fredrikson et al\cite{Fredrikson2015ModelInversion} (2015) produced model inversion attacks that enabled an adversary to exploit confidence values of predictions to infer sensitive information about individuals in the training dataset. They demonstrated that these attacks were successful in compromising regression-based pharmacogenetic models, decision trees deployed in machine-learning-as-a-service systems, and neural networks for facial recognition. It has also been seen that convolutional neural networks can memorise arbitrary labelings of the training data (Zhang et al\cite{Zhang2016rethinking}). More recently, Carlini et al\cite{carlini2021extracting} (2021) produced an attack that could expose the training data of a model by passing possible training points into the model and seeing if there is a strong indication of membership, indicated by low log perplexity of a point/term\footnote{In this context, given a generative sequence model $f_\theta$ and a sequence $\{x_i\}_{i\in[k]}$, the log perplexity is given by\[P_\theta(x_1,x_2,\ldots,x_k)=-\log_2\Pr[x_1,\ldots,x_k|f_\theta]=\sum_{i\in[k]}\left(-\log_2\Pr[x_i|f_\theta(x_1,\ldots,x_{i-1})]\right).\]}. That is, given a model, the log perplexities of possible training points are calculated and ranked in terms of how low each log perplexity value is. For example, in a not too large dataset, if an adversary wishes to find out if a particular sensitive value, like a social security number has been included, then the above ranking will indicate a lower log-perplexity for a sensitive value that is in the dataset over a random possible value of the sensitive feature. These attacks strongly demonstrate the necessity for private statistical learning.

\subsection{Differentially Private ERM}

\begin{table}[h]
\centering
\begin{tabular}{|l|c|c|c|}
\hline
\textbf{Type}          & \textbf{Introduced by} & \textbf{Year} & \textbf{Further Work} \\ \hline
Output Perturbation    &         \cite{chaudhurimonteleoni2008}               &     2008          &                 \cite{chaudhuri11a}   \\ \hline
Objective Perturbation &         \cite{chaudhurimonteleoni2008}               &     2008          &        \cite{chaudhuri11a}          \cite{kifer12privateconvexERM}  \cite{iyengar2019dpco} \cite{bassily2019privsco} \cite{Neel2019DifferentiallyPO} \\ \hline
Gradient Perturbation  &         \cite{williams2010probinference}               &      2010         &           \cite{song2013dpsgd}   \cite{bassily2014differentially}   \cite{Tran2021DPERMFair}   \cite{Xie2021DifferentialPS} \cite{wang2021dplis}  \\ \hline
\end{tabular}
\caption{Table summarising the different ways of performing differentially private stochastic gradient descent}
\label{tab:perturb}
\end{table}

 {Differentially private machine learning, abbreviated as DPML, however began as a field of study as early as 2008 with the introduction of privacy-preserving empirical risk minimisation (ERM) achieved via \emph{objective perturbation} by Chaudhuri and Monteleoni\cite{chaudhurimonteleoni2008}, which was further refined by Chaudhuri, Monteleoni, and Sarwate\cite{chaudhuri11a} along with the introduction of \emph{output perturbation} for private regularised ERM as an application of a result from Dwork, McSherry et al\cite{Dwork_McSherry_Nissim_Smith_2006} (2006).}

 {Consider the following setting: given a data space $\CX$ and a label set $\CY$, training data $\CD=\{(x_i,y_i)\in\CX\times\CY:i\in[n]\}$, and a loss function $\ell:\CY\times\CY\to[0,\infty)$, we wish to find a predictor $f:\CX\to\CY$ that performs well. Regularised ERM attempts to find such a well performing predictor $f$ by minimising the regularised empirical loss, given by}
 {\[J(f,\CD)=\frac1n\sum_{i=1}^n\ell(f(x_i),y_i)+\lambda N(f)\]}
 {where $\lambda\in[0,\infty)$ is a parameter for regularisation and $N$ is a real valued function that solely depends on $f$ and is not dependent on any $(x_i,y_i)\in\CD$ and is called the regularisation function that helps prevent overfitting.}

 {In the above setting, Chaudhuri et al\cite{chaudhuri11a} discuss a couple of ways of performing differentially private ERM for linear predictors $f$, $\CX=\R^d$ by an abuse of notation, they denote $f$ as a $d$-dimensional real vector and thus write $f(x)=f^Tx$.}
 {\subsubsection{Output Perturbation} This involves adding noise calibrated to the sensitivity of the regularised ERM to the output of the same. More precisely, given training data $\CD$, ERM outputs}
 {\[f_\text{priv}=\arg\min_f J(f,\CD)\]}
 {And then for a given $\eps>0$, a random noise vector $b$ is taken with respect to the probability density function $v(b)=\frac1\alpha e^{-\beta\Vert b\Vert}$, where $\alpha$ is a normalisation parameter and $\beta=\frac{n\lambda\eps}2$. Note that the pdf given here is that of the Gamma distribution resembles that of the Laplace distribution, and indeed, Chaudhuri et al show that for a given $\lambda$, the $L_2$-sensitivity of regularised ERM is upper bounded by $\frac 2{n\lambda}$, and that the following theorem holds.}

\begin{thm}
 {If $N$ is differentiable, $1$-strongly convex, and $\ell$ is convex and differentiable, with $|\ell'(z)|<1,\forall\,z$, then the aforementioned output perturbation method provides $\eps$-differential privacy.}
\end{thm}
 {\subsubsection{Objective Perturbation} This method was initially introduced by Chaudhuri and Monteleoni\cite{chaudhurimonteleoni2008} in 2008 in the context of logistic regression and reiterated in the general case of ERM in \cite{chaudhuri11a}. Objective perturbation as initially given in \cite{chaudhurimonteleoni2008} involves minimising a perturbed objective function which is given by}
\begin{equation}
 {J_\text{priv}(f,\CD)=J(f,\CD)+\frac 1n \langle b,f\rangle}\label{eq:obj-pert}
\end{equation}
 {Where $b$ is a random noise vector of dimension $d$ sampled with respect to the probability density function $v$ as given above but with $\beta=\frac\eps2$, and $\langle b,f\rangle:=b^Tf$.}

 {\cite{chaudhuri11a} presented a more sophisticated version of objective perturbation for ERM as given in algorithm \ref{objperturb}, with noise being added with respect to $\eps':=\eps-\log\left(1+\frac{2c}{n\lambda}+\frac{c^2}{n^2\lambda^2}\right)$ for a given privacy parameter $\eps>0$, and this choice of $\eps'$ is used by the authors to show that algorithm \ref{objperturb} is $\eps$-differentially private as in the following theorem.}
 {\begin{algorithm}
\caption{ERM with Objective Perturbation}\label{objperturb}
\begin{algorithmic}[1]
\REQUIRE Training data $\CD=\{z_i:=(x_i,y_i)\}$, privacy parameter $\eps$, regularisation parameter $\lambda$, parameter $c$
\STATE $\eps'\gets\eps-\log\left(1+\frac{2c}{n\lambda}+\frac{c^2}{n^2\lambda^2}\right)$.
\IF{$\eps'>0$}
    \STATE $\Delta\gets0$
\ELSE
    \STATE $\Delta\gets\frac{c}{n(\exp(\frac\eps4)-1)}-\lambda$
    \STATE $\eps'\gets\frac\eps2$
\ENDIF
\STATE Sample $b$ according to $v(b)=\frac1\alpha e^{-\frac{\eps'}2\Vert b\Vert}$
\STATE Compute $f_\text{priv}=\arg\min J_\text{priv}(f,\CD)+\frac12\Delta\Vert f\Vert^2$.
\end{algorithmic}
\end{algorithm}}

\begin{thm}
 {If $N$ is $1$-strongly convex and doubly differentiable, and $\ell$ is convex and doubly differentiable, with $|\ell'(z)|\leq 1$ and $|\ell''(z)|\leq c,\forall\,z$, then algorithm \ref{objperturb} is $\eps$-differentially private.}
\end{thm}
 {Note that these results only hold given that the loss and regularisation functions satisfy certain mathematical conditions.}

 {In 2012, Kifer, Smith, and Thakurta\cite{kifer12privateconvexERM} provided a better analysis for and slightly modified output perturbation to allow for similar privacy guarantees as the original version. They did this while relaxing the requirement of differentiability for the regulariser and with the addition of less noise, and expanded its scope of application to problems involving hard constraints. Loosely speaking, this involves minimising the perturbed objective function \ref{eq:obj-pert} while taking random noise $b$ from the Gamma distribution (similar to \cite{chaudhuri11a}) for $\eps$-differential privacy and from the Gaussian distribution for $(\eps,\delta)$-differential privacy.}

 Kifer et al also showed that where Chaudhuri et al's method provided an expected excess risk bound of $O\left(\frac{\zeta\Vert\hat\theta\Vert_2 p\log p}{\eps \sqrt n}\right)$, where $p$ is the dimension of the input data, their method provided a better bound of $O\left(\frac{\zeta\Vert\hat\theta\Vert_2 \sqrt{p\log (1/\delta)}}{\eps \sqrt n}\right)$. Later, Jain and Thakurta\cite{jain14neardimensionindependent} (2014) provided techniques to perform output and objective perturbation by drawing noise only from the Gaussian distribution, and not from the Gamma distribution as in \cite{chaudhuri11a}, that have expected excess risk bounds independent of the dimension $p$. More precisely, Jain and Thakurta's technique achieved an excess risk bound of $O\left(\frac{(\log^2 n)(\zeta)^2\Vert\theta\Vert_2\sqrt{\log(1/\delta)+\eps}}{\eps\sqrt n}\right)$. Duchi et al\cite{duchi2013minimax} (2013) provided a formal minimax risk based framework for local differential privacy on statistical estimators, and provided tight bounds on the expected excess risk in locally differentially private convex risk minimisation, and a gradient perturbed and locally differentially private version of stochastic gradient descent that achieves these bounds.

 {However these approaches impose the requirement that an exact optimum is arrived at for these guarantees to hold. This is often not possible in a practical setting due to various issues, such as those involving numerical precision in computers and the iterative nature of most optimisers in practice. This leaves these algorithms open to attacks in a practical setting. A prominent illustration is given in \cite{mironov2012lsbs} by Ilya Mironov showing that practical implementations of something as basic as the Laplace mechanism is vulnerable to attacks due to irregularities of floating-point implementations of the mechanism, and this vulnerability is inevitably carried over to differentially private ERM.}

 {More recent works on private stochastic convex optimisation by Iyengar et al\cite{iyengar2019dpco} (2019) and Bassily et al\cite{bassily2019privsco} (2019) do not require convergence to an exact minimum. Works like these show that it suffices to obtain an approximate minimum for the objective function which makes these forms of objective perturbation more feasible in a practical setting for stochastic convex optimisation. However, these papers still impose the condition of convexity on the loss function. Neel et al (2019)\cite{Neel2019DifferentiallyPO} does away with the requirement for convexity of the loss function, instead merely requiring it to be bounded while working with a discrete domain. For a continuous domain, the authors merely in addition that the loss function be Lipschitz in its continuous parameter.}
 
 {\subsubsection{Gradient Perturbation} Another popular way of performing differentially private machine learning is via \emph{gradient perturbation} which involves performing gradient descent with noise addition. The idea of performing noisy gradient descent was initially proposed by Williams and McSherry\cite{williams2010probinference} in 2010. A simple version of gradient descent was proposed by Song, Chaudhuri, and Sarwate\cite{song2013dpsgd} in 2013, which involved performing stochastic gradient descent (SGD), w.r.t. a convex loss function and an $L_2$-regularised objective (ergo strongly convex loss functions), with the following SGD update iteration,}
 {\[w_{t+1}=w_t-\eta_t(\lambda w_t+\nabla\ell(w_t,x_t,y_t)+Z_t)\text{ for }Z_t\sim\CD\]}
 {where $\CD$ is a distribution with the probability density function $\rho(z)=e^{\left(\frac\alpha2\right)\Vert z\Vert}$, and $Z_t$ is some random noise drawn from $\CD$.}
This guarantees $\eps$-DP given that the norm of the gradient of the loss function, $\nabla\ell(w_t, x_t, y_t)\leq 1,\forall\,w$, and $\forall\,(x_t,y_t)$.

Bassily et al\cite{bassily2014differentially} (2014) provided improvements to gradient perturbation, with the requirement that the loss function is Lipschitz bounded and that the domain of optimisation is bounded. Their algorithm adds Gaussian noise to the computed gradient, and thus via advanced composition guarantees achieves better and tighter risk bounds (vis-\`a-vis previous works) that are fairly comparable to theoretical bounds. They also provided algorithms for tasks like stochastic gradient descent, exponential sampling based convex optimisation etc. Their popular algorithm for private SGD is given by algorithm \ref{gradperturbbassily}.\footnote{$\sim_u$ denotes choosing uniformly at random, and $\mathbb{I}_p$ is the $p$-dimensional identity matrix.}

\begin{algorithm}
\caption{Differentially Private Stochastic Gradient Descent}\label{gradperturbbassily}
\begin{algorithmic}[1]
\REQUIRE Training data $\CD=\{d_1,\ldots,d_n\}$, privacy parameters $(\eps,\delta)$, $L$-Lipschitz loss function $\ell$, convex set $\CC$, and the learning rate function $\eta:[n^2]\to\R$.
\STATE Arbitrarily choose any $w_1$ from $\CC$.
\FOR{$t = 1$ to $n^2-1$}
    \STATE Pick $d\sim_u\CD$ with replacement.
    \STATE $w_{t+1}=\prod_C\left(w_t-\eta(t)\left[n\nabla\ell(w_t;d)+b_t\right]\right)$ where $b_t\sim\CN(0,\mathbb{I}_p\sigma^2).$
\ENDFOR
\STATE Output $w^\text{priv}=w_{n^2}$.
\end{algorithmic}
\end{algorithm}

The expected excess risk for algorithm \ref{gradperturbbassily} is shown to be $\tilde O\left(\frac{\Vert\CC\Vert_2L\sqrt{p\log(1/\delta)}}{\eps}\right)$, and in addition, it satisfies $(\eps,\delta)$-differential privacy.

Abadi et al\cite{Abadi2016DeepLearnDP} (2016) provided a gradient perturbation algorithm named DP-SGD for deep learning purposes. This is discussed in some more detail in section \ref{sec42}.

In 2016, Papernot et al provided an algorithm for gradient perturbed SGD that clips the gradients so as to bound their norms to allow noise addition via sensitivity based mechanisms, even for non-Lipschitz loss functions with unbounded variants.


\subsubsection{Improvements, Practical Issues and Mitigation} Differentially private gradient descent comes with a set of auxiliary challenges that have been the subject of study to make it more practically feasible.

In 2021, Tran et al\cite{Tran2021DPERMFair} showed that differentially private ERM, in particular gradient perturbation involving gradient clipping, can incur a higher level of unfairness towards certain vulnerable/minority groups than non-private ERM, and they provide a mitigating algorithm for differentially private ERM that corrects for better fairness and higher utility after noise addition.

Xie et al\cite{Xie2021DifferentialPS} (2021) note that intuitively the value of the gradient, and hence that of the gradient norm is inversely proportional to the number of iterations completed leading to varying privacy leakage risk across iterations, and that most approaches to implementing DP-SGD involve splitting the privacy budget evenly across iterations. Also, as the training process approaches convergence, the values of the gradients, being small, must be reported more accurately. They propose an adaptive, noise-reducing algorithm for DP-SGD that involves adaptively allocating a share of the privacy budget to each iteration.

Chen et al\cite{Chen2020UnderstandingGC} (2020) examine DP-SGD from a geometric perspective and note that gradient clipping can lead to a substantial bias in the update direction in each step of training, and may even lead to the update leading away from the optimum in some cases. They provide theoretical and empirical analyses in this regard and present a correction method to reduce the aforementioned bias by adding noise to the gradient prior to clipping.

Liu and Talwar\cite{liu2018private} showed in 2018 that repeated hyperparameter selection for running ML models multiple times while finetuning hyperparameters can increase privacy loss significantly as opposed to a single run of an ML algorithm. In addition, they proposed a mitigating strategy that involves searching for hyperparameters randomly along with a random stopping rule. Papernot and Steinke\cite{papernot2021hyperparameter} built upon this work in 2021, analysing this problem via the lens of R\'enyi DP and demonstrated this additional privacy loss issue successfully for SVMs trained on certain data distributions. They improved upon the mitigating strategy, providing improvements to the stopping rule that further reduced privacy loss significantly. Chaudhuri et al\cite{chaudhurivinterbo2013stabilitybasedvalidation} (2013) pointed out that validating a model and training it with different training parameters leads to an increase in privacy loss (training a model on the same set of $\eps$-DP perturbed training data $k$ times yields $k\eps$-DP), and thus propose an approach for carrying out this validation exercise without splitting the privacy budget or the training set across rounds of training. They define stability conditions for the validation score function over changes in the training and validation sets with some privacy parameters $\eps,\delta$, and use that to produce sufficient conditions for differentially private guarantees on the validation procedure.

In practice, performing per-example gradient clipping a na\"ive implementation of DP-SGD incurs takes much more time for private gradient descent over its non-private counterparts when using commonly available deep learning frameworks like PyTorch which only provide the aggregated gradient for each batch using auto-differentiation. Lee and Kifer (2020)\cite{Lee2020Scaling} remedy this by introducing new methods for per-example gradient clipping that is compatible with  auto-differentiation in these frameworks, and thus provide a much faster practical implementation of DP-SGD. They achieve this by extending a trick given by Goodfellow\cite{Goodfellow2015EfficientPG} for calculating per-example gradients using auto-differentiation to various neural networks: given the preactivation of a layer, $z=Wx+b$, the auto-differentiator is asked to calculate $\frac{\partial L}{\partial z}$ and computing the per-example gradient as $\frac{\partial L}{\partial W}=\frac{\partial L}{\partial z}\otimes x$ for the example $x$.\footnote{$\otimes$ here is the vector outer product.} Lee and Kifer extend this to neural networks and use it to compute the per-example gradient norms with the auto-differentiator and subsequently the clipping weights $v_i=\min(1,\frac{C}{\nabla_\theta\ell(f_\theta(x_i),y_i)})$ for a machine learning model $f_\theta$ with parameters $\theta$ and a clipping bound $C$. The auto-differentiator is then asked to provide the gradient of the reweighted objective function $L=\sum_{x_i\in B}v_i\nabla_\theta\ell(f_\theta(x_i),y_i)$ and then Gaussian noise can be added to the resulting value to privatise it.

Wang et al\cite{wang2021dplis} (2021) demonstrated that differentially private machine learning algorithms like DP-SGD can affect the prediction accuracy of the resulting privately trained model and can be highly unstable as different runs may yield models with significantly different prediction accuracies. This is due to the loss function having irregularities and several local minima, and perturbing the gradients can lead the algorithm down a different path than in a non-private setting. They thus propose smoothing the loss function so it has one, flat loss surface, and thus the training will be more robust and tolerant to noise addition.

Tr\`amer and Boneh (2021)\cite{Tramer2021Vision} study how, ceteris paribus, varying the batch size and the learning rate jointly affect the learning of models like private image classifiers, and they propose a linear scaling rule that states that upon scaling the batch size and the learning rate by the same constant yields models with the same performance. They also note that on moderate privacy budgets, simpler linear models trained on handcrafted features outperform end-to-end deep learning algorithms on several tasks even if the latter may have more trainable parameters. To outperform these linear models trained on handcrafted features, these private deep learning models require either much more private data, or access to features learned on public data from a similar domain.

Ligett et al\cite{ligett2017accuracy} noted that most approaches to making empirical risk minimisation differentially private focus on fixing the privacy parameters (viz. $\eps$) first and then attempt to maximise the accuracy of the learning subject to that. They introduce a noise reduction framework for differentially private ERM that with respect to a specified accuracy constraint searches the space of privacy levels that empirically satisfies the accuracy constraint. This search however can be computationally expensive as it requires running the learning algorithm for various privacy levels and whether they satisfy the accuracy constraint empirically.

Applying techniques like DP-SGD on a large-scale, as in large neural networks, continues to be a practical challenge. Da et al \cite{da2021largescale} (2021) introduce a method to make it more feasible by reducing memory costs by modifying how relevant weight and gradient vectors/matrices are represented and presenting a correspondingly modified gradient perturbation algorithm.
\paragraph{Private Aggregation of Teacher Ensembles (PATE)}

\begin{figure}
    \centering
    \includegraphics[width=\textwidth]{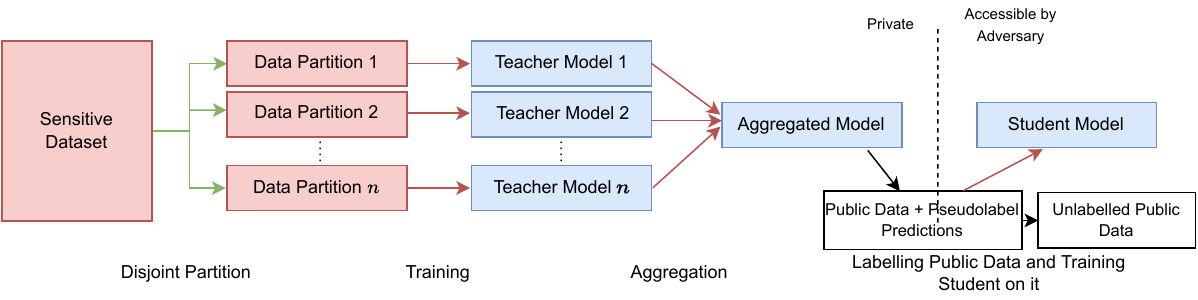}
    \caption{Diagram illustrating the working of PATE \cite{Papernot2018PATEScalable}}
    \label{fig:PATE}
\end{figure}
Taking inspiration from the sample-and-aggregate framework introduced by Nissim et al in \cite{Nissim2007SmoothSens} and the concept of distillation of models as a strategy to protect against adversarial input perturbation/poisoning attacks\cite{papernot2015distillation}, Papernot et al\cite{papernot2018knowledgetransfer} (2016) introduced a technique for private knowledge transfer from private teacher models to a student model by aggregating votes from the teacher models and adding random noise to the aggregation process.

This was further improved on by Papernot et al\cite{Papernot2018PATEScalable} in 2018.\footnote{Not to be confused with the 2016 paper coauthored by Papernot introducing PATE.} PATE (described in Figure \ref{fig:PATE} is implemented by choosing a number of $k$ many disjoint teacher models and partitioning a sensitive dataset into disjoint subsets, one each to train a private teacher model on, and then they are given public unlabelled data and they output (pseudo-)labels, and these are treated as \emph{votes} that are aggregated into a vote histogram to which noise, viz. Laplace noise, is added and then the prediction with the highest number of votes in the noised histogram is released. This is used to train the student model. Ideally since these teacher models are disjoint, i.e. trained on disjoint subsets of the training data, and are assumed to have high accuracy, then a overwhelming majority of them shall vote for one (hopefully correct) label. In this case, the most voted for prediction shall have a large difference of votes from its nearest competitors and thus this can be released exactly, else an output with some randomisation is output by the ensemble. Then the student model is trained using these private labels and public data, using a fixed number of queries as the privacy cost increases with the number of queries, in a semi-supervised manner.

The case for PATE's privacy is further helped along by the fact that any given training point can belong to at most one teacher model's training dataset and given that the teacher models, even those without that point in their respective training datasets, almost always predict correctly, thus intuitively providing a measure of differential privacy. One noteworthy takeaway is that PATE is model-agnostic and can be implemented atop any chosen type of teacher models.

\subsection{Applications in Deep Learning}\label{sec42}

Table \ref{tab:sec42} provides a brief summary of the works discussed in this section.

\begin{table}[!h]
\begin{tabular}{|l|c|p{0.5\linewidth}|}
\hline
\textbf{Category}                                       & \textbf{Work} & \textbf{Remarks} \\ \hline
\multirow{2}{*}{Gradient Perturbation in Deep Learning} &    \cite{Abadi2016DeepLearnDP}           &             Introduces DP-SGD     \\ \cline{2-3} 
                                                        &    \cite{Bu2020DeepLW}        &  Extends \cite{Abadi2016DeepLearnDP} with the use of Gaussian DP               \\ \hline
\multirow{4}{*}{DP for Computer Vision}                 &  \cite{Huang2019DPCNN}             &          Training CNNs with DP (DPAGD-CNN)        \\ \cline{2-3} 
                                                        &       \cite{Zhu_2020_CVPR}        &     Private-$k$NN             \\ \cline{2-3} 
                                                        &     \cite{Luo_2021_CVPR}          &   Transfer learning for computer vision using sparse subnetworks and DP               \\ \cline{2-3} 
                                                        &       \cite{Golatkar_2022_CVPR}        &  Pre-training computer vision models on public data with private finetuning (AdaMix)                \\ \hline
\multirow{5}{*}{DP for Graph Learning Tasks}            &     \cite{sajadmanesh2021gnn}          &         Node-level privacy for GNNs using randomised response/LDP         \\ \cline{2-3} 
                                                        &      \cite{Olatunji2021GNNDP}         &  Protection of private GNNs by training public student models                \\ \cline{2-3} 
                                                        &    \cite{Sisong2021SGX}           &   Attempt to convert LDP based node privacy approaches into a CDP setting               \\ \cline{2-3} 
                                                        &        \cite{Mueller2022DifferentiallyPG}       &         Adapt DP-SGD to train GNNs         \\ \cline{2-3} 
                                                        &  \cite{bun2021differentially}             &       Efficient correlation clustering on graphs using DP           \\ \hline
\multirow{5}{*}{DP for Natural Language Processing}     &        \cite{Feyisetan2020PrivacyAU}       &           Language Modelling with DP using metric-based LDP       \\ \cline{2-3} 
                                                        &     \cite{Fernandes2019GeneralisedDP}          &        Language Modelling with DP using metric-based LDP          \\ \cline{2-3} 
                                                        &        \cite{mcmahan2018dplearning}       &      Private LSTM using private federated averaging            \\ \cline{2-3} 
                                                        &      \cite{Li2021LargeLM}         &
                                              Efficient finetuning of large transformer models using DP-SGD          \\ \cline{2-3}
                                        &       \cite{Kerrigan2020DifferentiallyPL}        & Privately finetuning an initial public language model for better accuracy
                                                        \\ \hline
\end{tabular}
\caption{Summary of some of the applications of differential privacy to various (deep) learning tasks.}
\label{tab:sec42}
\end{table}

\subsubsection{Foundations of Differentially Private Deep Learning}
The study of applying differential privacy to deep learning models viz. neural networks is a natural extension of the work done in differentially private machine learning. This was initiated by Abadi, Chu et al in 2016\cite{Abadi2016DeepLearnDP}, which adapted existing gradient perturbation techniques from works like \cite{bassily2014differentially} to create an algorithm called DP-SGD for differentially private training of neural networks. Abadi et al propose sampling a subset $\tilde\CD$ of a fixed size $L$ of a dataset $\CD$ of size $n$ uniformly at random and then for every $z\in\tilde\CD$, the gradient of the loss function is calculated and clipped to have an $\ell_2$ upper bounded by some parameter $U$, following which they are averaged and privatised by the addition of Gaussian noise calibrated to the sensitivity bound $U$, and this is performed iteratively until the termination of the training process. Note that due to clipping, the loss function is not restricted to being Lipschitz. This procedure endows $(\eps,\delta)$-differential privacy.

Inspired by the introduction of $f$-DP, particularly Gaussian DP, Bu et al\cite{Bu2020DeepLW} (2020) extend the work done by Abadi et al\cite{Abadi2016DeepLearnDP} by using Gaussian DP. The authors provide an improved analysis of differentially private deep learning, and noisy versions of stochastic gradient descent and Adam optimisation using Gaussian DP and its benefits over previously defined variants, such as improved handling of composition and subsampling, without requiring the development of sophisticated analysis tools (viz. the moments accountant) as in \cite{Abadi2016DeepLearnDP}. It is shown that similar privacy guarantees can be achieved by the use of $f$-DP/Gaussian DP as compared to using $(\eps,\delta)$-DP or the moments accountant, thus yielding models with higher utility.

\subsubsection{Applications of Differentially Private Learning to Particular Tasks}
\paragraph{For Computer Vision}
In this subsection, we describe a few applications of differential privacy to deep learning tasks.
There has been some work on applying differential privacy to deep learning models for computer vision (viz. convolutional neural networks (CNNs)). Huang et al\cite{Huang2019DPCNN} (2019) present an algorithm which they call DPAGD-CNN (Differentially Private Adaptive Gradient Descent for CNNs) which trains CNNs by varying the amount of privacy budget available for adding noise to the gradient and optimal step size adaptively and accounting for privacy using zCDP.

For tasks with limited data availability like those related to computer vision, getting ample labelled data is often expensive and splitting the datasets to train disjoint teacher models as in PATE will yield suboptimal accuracy. Zhu et al\cite{Zhu_2020_CVPR} (2020) proposed a method called \emph{Private-$k$NN} which avoids splitting the training private dataset, given a student model, a feature extractor, and public unlabelled data. It involves picking a random subset from the private dataset with Poisson sampling, and then running the $k$NN algorithm on it with the aid of the feature extractor, and this process is performed iteratively, with the feature extractor being updated by the student (deep) model with every iteration. Query responses are released given that they pass noisy screening by having a large degree of consensus in voting. Subsequently, the student model is trained  using the released query responses as pseudo-labels in a self-supervised manner. The authors demonstrate, with R\'enyi DP privacy accounting and by the principle of privacy-amplification-via-subsampling, that their method provides significant improvements on existing methods' privacy bounds despite its iterative nature, rendering it a practical private deep-learning method for computer vision.

Luo et al\cite{Luo_2021_CVPR} (2021) note that the assumption on the availability of ample public data made by most DP transfer learning models can be unrealistic, especially for computer vision and visual recognition tasks, and that traditional models of performing computer vision tasks with differential privacy (viz. DP SGD) work only on simple datasets and shallow networks. They contend that in order to improve the privacy-utility tradeoff in this context, the number of training parameters must be minimised. To that end, they provide novel methods of performing transfer learning that produce an optimal, sparse subnetwork. 

Golatkar et al\cite{Golatkar_2022_CVPR} (2022) note that pre-training models on public data might be beneficial for language models, but for computer vision tasks, they can lead to a heavy privacy-utility tradeoff. To this end, they introduce AdaMix which involves pre-training a model with few-shot or cross model zero-shot learning on public data prior to private finetuning of the model using noised, projected, private gradients (w.r.t. an adaptively changing clipping threshold that is large at first and reduces in size to ensure higher accuracy at first and better privacy towards the end of the training) using a private dataset, which vastly improves on the privacy-utility tradeoff of its baselines.

\paragraph{For Graph Neural Networks and Learning on Graphs} 
There has been a brief body of work regarding differential privacy for graph neural networks (GNNs), which are neural networks that work on graph based data that comprises of several nodes that contain some node data and are joined by edges according to how they are related in a given context. Several of those, prominently Sajadmanesh et al\cite{sajadmanesh2021gnn} (2021), propose using locally differentially private techniques like randomised response to perturb (randomly chosen) bits of the node data at the node level prior to training to ensure privacy of the nodes' sensitive data. Others like Olatunji et al (2021) \cite{Olatunji2021GNNDP} focus on protecting proprietary/sensitive GNN models by using the secret GNN model as a teacher to teach a public \lq\lq student" model without revealing the private model's weights. Olatunji et al achieve this with the use of central differentially private mechanisms along with privacy-amplification-via-subsampling by randomly selecting an induced subgraph of the teacher model's graph and using it to train the student model. 

Sisong et al\cite{Sisong2021SGX} (2021) aim to convert the LDP based node privacy problem into a centrally differentially private one by using (seemingly) trusted secure hardware like Intel's SGX to do differentially private calculations prior to releasing node data to the analyst for training the GNN, thus leading to significant improvements in terms of utility and accuracy of training. However Intel's SGX has been shown to be vulnerable to various attacks\cite{Fei2021SGX} and has been deprecated, rendering this approach ineffective in a practical setting for the moment.

More recently, Mueller et al\cite{Mueller2022DifferentiallyPG} (2022) presented a method to perform graph level classification on multi-graph datasets by adapting the DP-SGD algorithm to use it to train graph neural networks.

Eli\'a\v s et al\cite{elias2020syntheticgraphs} (2020) proposed a $(\eps,\delta)$-differentially private algorithm to produce, given a graph $G$ (with potentially sensitive information) with $n$ vertices and $m$ edges, a synthetic graph $G'$ that approximates all the cuts of $G$ up to an additive error of $O\left(\sqrt{\frac{mn}\eps}\log^2(\frac n\delta)\right)$, and $o(m), \forall\,m>n\log^C n$, providing good approximations for sparse graphs as well. Using ideas from this and differentially private noise addition, Bun, Eli\'a\v s, and Kulkarni\cite{bun2021differentially} (2021) proposed efficient differentially private algorithms for performing privacy-preserving correlation clustering for weighted and unweighted graphs with subquadratic error.\footnote{A task in unsupervised machine learning that involves clustering a set of objects based on the given information about how similar/dissimilar an object is to another. Introduced by Bansal, Blum, and Chawla\cite{Bansal2002correlation} in 2002.}

\paragraph{For Natural Language Processing (NLP)}
There has also been some work on differentially private NLP/language modelling. Using DP-SGD from \cite{Abadi2016DeepLearnDP} na\"ively can lead to punishing privacy-accuracy tradeoffs for language modelling. Feyisetan et al\cite{Feyisetan2020PrivacyAU} (2020) and Fernandes et al \cite{Fernandes2019GeneralisedDP} (2019) instead employ differential privacy for language modelling, for text perturbation to ensure geo-indistinguishability in location data and for author obfuscation respectively, with respect to metric-based relaxations of local DP, and add noise to the vector embedding of a word.

McMahan et al\cite{mcmahan2018dplearning} (2018) provide a private LSTM model that utilises a noised version of the federated averaging algorithm\cite{McMahan2016FL}\footnote{Federated averaging shall be discussed in the next subsection.} to give a trained model with strong privacy guarantees and a high accuracy relative to its non-private counterpart for a large enough dataset. 

Kerrigan et al\cite{Kerrigan2020DifferentiallyPL} (2020) improve on the state-of-the-art differentially private language models in terms of the privacy-utility tradeoff by first training a public/non-private model on a large public dataset before private finetuning by using DP-SGD using private, out-of-distribution dataset.

Li et al\cite{Li2021LargeLM} (2021) describe methods to efficiently fine tune large transformer models with millions of parameters directly with $\eps$-differential privacy, using DP-SGD, for $\eps\in\{3,8\}$. Prominently, they introduce a relatively computationally efficient and memory efficient technique known as \emph{ghost-clipping} to improve upon the utility of DP-SGD, which when na\"ively implemented incurs substantial memory overhead while clipping per-example gradients. Ghost clipping involves arriving at the per-example gradient norms without substantiating the per-example gradients themselves, and achieves a significantly lower memory complexity as a result. This is inspired by the technique to efficiently calculate gradient norms introduced by Goodfellow in 2015\cite{Goodfellow2015EfficientPG}.

\subsection{In Federated Learning}


When it comes to privacy-preserving machine learning, federated learning is often discussed along with DPML. Consider a scenario with multiple users that hold personal data that is required for training/updating a model by a central authority/server, but the users may not want to part with their personal, and potentially sensitive, data in a raw form. Apart from privacy concerns, sending huge volumes of one's personal data, viz. photographs or voice recordings, will be expensive in terms of communication costs. This is a privacy preserving variant of an older technique known as \emph{distributed (machine) learning} which involves outsourcing certain training tasks for a model to multiple nodes which are all under the control of the same central authority.

\begin{figure}
    \centering
    \includegraphics[width=0.8\textwidth]{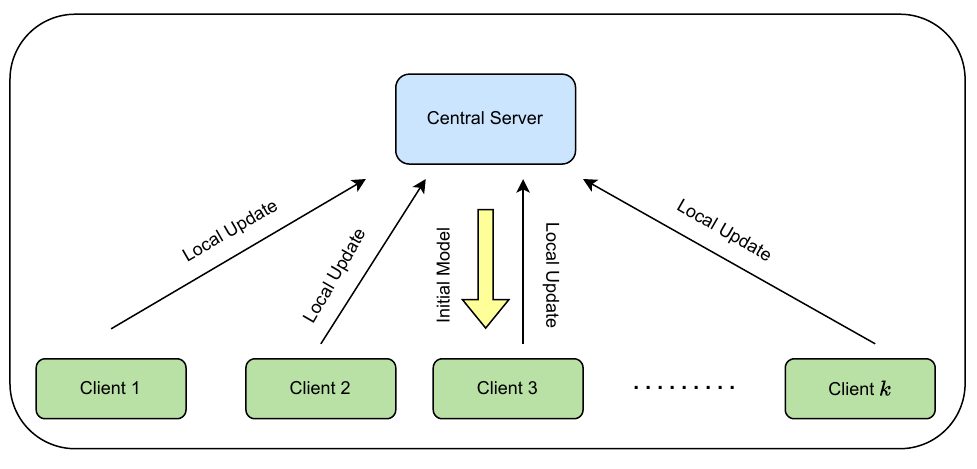}
    \caption{An Overview of the Federated Learning Process. The central server provides the clients with a global, initial model. The clients train the model on their local data and send the result of their local training (local updates) back to the server. The server then aggregates the local updates and updates the global model.}
    \label{fig:FL}
\end{figure}

Introduced by McMahan et al\cite{McMahan2016FL} in 2016, federated learning, or FL for short, provides a solution for this by having the central server provide the devices with an initial model and have them train said model on their personal data, and send the resulting weight updates or gradients to the central server to be aggregated as a weighted arithmetic mean, via an algorithm known as \texttt{FederatedAveraging}. This process has been shown to yield pretty accurate models as a result. Figure \ref{fig:FL} provides an overview of the process of federated learning.

Federated learning is a powerful privacy-preserving and communication cost-cutting technique, but it has a variety of facets that have been worked on since its conception; we shall restrict our focus to privacy-related concerns and application of DP to FL in particular. Attacks on the models resulting from FL, including membership inference attacks as discussed earlier, cannot be ruled out, and one significant line of work that seeks to improve on the privacy of FL is by applying differential privacy to it.

Works like those by Wei et al (2020)\cite{Wei2020NbAFL} use a commonly used method of clipping the weight updates with respect to a specified clipping bound $C$ and adding differentially private random noise to them (viz. by applying the Gaussian mechanism) to achieve DP guarantees. Paul et al\cite{paul2020flaps} introduced FLaPS, a paradigm to conduct federated learning with scalability and enhanced privacy guarantees with differential privacy guarantees being endowed by BUDS\cite{Sengupta2020BUDS} and ARA\cite{Paul2020BUDS}
; FLaPS involves taking devices participating in the training process and clustering them into silos and assigning a cluster centre among the silo members, and aggregating their privatised data securely using ARA and training the initial model provided by the central server using the aggregated data and privatising the model's weights with BUDS. Following this, the cluster centres send these reports to the central server for aggregation using ARA and \texttt{FederatedAveraging}, in that order. These methods are shown to provide DP guarantees, and the latter additionally provides scalability and communication efficiency by reducing the number of links between individual devices and the central server.

Truex et al\cite{truex2020ldpfed} (2020) present LDP-Fed which performs federated learning with local differentially privatised client updates that are accepted or rejected uniformly at random (thus achieving privacy loss reduction by privacy-amplification-via-subsampling) before the central server aggregates the accepted updates.

Hu et al\cite{hu2020concentrated} (2020) have the members of a set of participating clients $\Omega^t$ train their local models with gradient perturbation with Gaussian noise, and then perturb their respective local weight updates $p_i^t$ by adding some random value $r_i^t$ to it such that $\sum_{i\in\Omega^t}r_i^t=0$, generated using a protocol involving a certain pseudorandom function and a seed agreed upon by participating and mutually communicating clients during each round. This allows the central server, which may be honest-but-curious, to aggregate the local weight updates securely without having an idea as to what each of these local weight updates actually look like.

Girgis et al\cite{girgis21apmlr} (2021) presented a method to learn a model with communication constraints and provide privacy guarantees with reasonable utility of the model. The server chooses a random subset of clients at each round, each of which use a random subset of their personal training data, and privatise their responses by clipping their gradients and using an LDP mechanism to privatise their gradients. Following this, the aggregating central server receives a random permutation of these updates after shuffling via a secure shuffler. This essentially is a subsampled shuffle model. The same authors \cite{girgis2021renyi} (2021) extend their work on CLDP-SGD to present an analogous differentially private approach to distributed learning, and provide a stronger privacy analysis of CLDP-SGD using RDP.

Andrew et al\cite{andrew2021differentially} (2021) note that there is no a priori optimal value for the update clipping bound for noise addition across various learning tasks, and the update norm distribution is dependent on the model, client learning rate, amount of data possessed by each client and other such parameter. The authors then propose adaptively choosing a clipping bound at a particular quantile of the update norm distribution at any point in time, instead of adhering to a fixed clipping bound specified beforehand, thus producing a method that is shown to outperform any prior choice of a fixed bound.

Truex et al\cite{Truex2019AHA} propose a hybrid approach to federated learning which combines differential privacy with secure multiparty computation (SMC) with a tunable trust parameter to provide better model accuracy along with provable privacy guarantees and protecting against extraction attacks and collusion threats. The clients are queried by the server/aggregated and respond with adding differentially private noise to their query responses and encrypting them homomorphically, which the aggregator can then aggregate. The aggregator then asks a sufficiently large subset of clients (determined by the trust parameter) to help decrypt the aggregate value. The combined usage of differential privacy and SMC makes sure that the model output is differentially private and that the exchange of non-private messages remains protected without information leaks. The authors also provide algorithms to implement this approach on various ML and DL models viz. CNNs, SVMs, decision trees, etc.

Papernot et al\cite{Papernot2021TemperedSA} (2021) note that the implementation of differentially private techniques for learning are often used on models that are shown to be successful in a non-private setting, leading to sub-optimal private training (sub-optimal utility for a given privacy level) in some cases. They therefore suggest selecting model architectures explicitly for private training in the first place. They also note that the choice of activation function (and the bounds on them or lack thereof) plays a major role in determining the sensitivity of private deep learning, and that bounded activation functions like the tempered sigmoid function consistently outperform unbounded activation functions like ReLU. 

Xu et al\cite{Xu2021ADP2SGDAD} (2021) discuss the application of differential privacy to asynchronous decentralised distributed learning and introduce A(DP$)^2$SGD, which is a differentially private version of asynchronous decentralized parallel stochastic gradient descent (ADPSGD), which helps protect from information leaks during communication between nodes. This essentially entails the addition of Gaussian noise to the local gradient by each client while carrying out the ADPSGD process.
\section{Industrial Deployments}

\begin{table}[h]
\centering
\begin{tabular}{|l|l|l|l|}
\hline
\textbf{Organisation}      & \textbf{Work/Industrial Implementation} & \textbf{Year}  \\ \hline
\multirow{4}{*}{Google}    & RAPPOR                                  & 2014                           \\ \cline{2-3} 
                           & Privacy on Beam                         & 2021                            \\ \cline{2-3} 
                           & DP SQL                                  & 2020                            \\ \cline{2-3} 
                           & Plume                                   & 2022                            \\ \hline
Apple                      & Sequence Fragment Puzzle                & 2017                         \\ \hline
\multirow{3}{*}{Microsoft} & PINQ                                    & 2010                            \\ \cline{2-3} 
                           & One-Bit Estimation                      & 2017                            \\ \cline{2-3} 
                           & LinkedIn Audience Engagement API        & 2020                            \\ \hline
Uber                       & FLEX                                    & 2018                           \\ \hline
\end{tabular}
\caption{Table summarising the different practical deployments and industrial implementations of DP discussed in this section.}
\label{tab:deploy}
\end{table}

The utility of differential privacy has been widely recognised by industry and data handling organisations, but practically implementing it in a manner that is easy to use, even by users without an in-depth understanding of differential privacy has proven to be a challenge and an important task for its widescale adoption. 
We shall very briefly discuss some practical implementations (mentioned in table \ref{tab:deploy}) of DP below.
\subsection{By Google}
\paragraph{RAPPOR}RAPPOR\cite{RAPPOR} (2014), as discussed, is the first well-known industrial implementation of differential privacy, which was deployed by Google for the Chromium browser. It consists of a few layers: a bloom filter and then two rounds of randomised response (for logitudinal privacy and then to deidentify the user from the bloom filter output with one application of randomised response). A string from a known universe of strings is passed through these layers by a user, following which the resulting reports are sent from each user to an aggregator who infers useful information from the resulting aggregation.
\paragraph{Privacy on Beam and Differentially Private SQL}Google introduced an end-to-end differential privacy solution for Apache Beam called Privacy on Beam\cite{GooglePrivonBeam} that can be used without any particular expertise with differential privacy.
Wilson et al\cite{Wilson2020DifferentiallyPS} (Google, 2020) introduced a system to answer various SQL queries with user-level differential privacy, and empirically demonstrate the utility, robustness, and scalabity of this system.
\paragraph{Plume (2022)} Amin et al\cite{amin2022plume} adapt and modify the MapReduce model\cite{MapReduceDean} of distributed computation to introduce Plume for Google, which provides scalable differential privacy for large databases. The privacy budget is controlled by limiting how many keys any user can contribute records to, then out of these keys a safe key set $S$ is produced, and instead of using a non-private aggregation algorithm as in MapReduce, a differentially private mechanism like the Laplace mechanism is used to add noise to the values corresponding to the keys in $S$. 
\subsection{By Apple}
Apple\cite{AppleDP} took inspiration from the Count Sketch algorithm which was developed by Charikar et al\cite{Charikar2002CountSketch} to efficiently estimate the most frequent items in a data stream using limited storage space. 

Apple's privacy system in the paper utilised an LDP randomisation technique known as Sequence Fragment Puzzle for privatisation at the user-level. Each word is broken up into fragments and the frequency of each word is calculated. The user then concatenates a random substring of the string (word) with the hash of the entire string and privatises it, and transmits it with the index at which the substring starts in the string. The transmission of these messages is further endowed with privacy and security guarantees by delaying the transmission of these messages, then randomly subsampling the messages that are received and removing identifying details like the user's IP address from the messages and using TLS encryption to send it to the server.

This work notably improves on RAPPOR in that while RAPPOR only supports the privatisation of the members of a fixed universe of strings, Sequence Fragment Puzzle allows for the discovery of new strings. However, it has faced criticism about some of its facets from works like \cite{Tang2017PrivacyLI}.
\subsection{By Microsoft}
\paragraph{PINQ}McSherry\cite{McSherryPINQ} (2010) introduced Privacy Integrated Queries (PINQ), an API resembling and extending Language Integrated Queries\footnote{Which is an SQL-like declarative query language extension for .NET languages} (LINQ), which can be used to perform privacy-preserving data analysis on sensitive datasets. Proserpio et al\cite{wPINQ} designed an extension to PINQ known as Weighted PINQ or wPINQ that assigns weights to every row in the database and then scales the weights of a row in a join to ensure that the overall sensitivity is 1. It supports general equijoins.

\paragraph{One-Bit Estimation} Ding et al\cite{DingMicrosoft} (2017) utilise randomised response to generate local reports by users starting from a raw local value $X_i\in[0,m]$ as follows,
$$Y_i=\begin{cases}1\text{ with probability}=\frac1{e^\varepsilon+1}+\frac{X_i}m\cdot\frac{e^\varepsilon-1}{e^\varepsilon+1}\\0\text{ otherwise}\end{cases}.$$
These are then aggregated to gain an unbiased average report from the local reports as follows,
$$\hat\mu=\frac mn\sum_{i=1}^n\frac{Y_i\cdot(e^\varepsilon+1)-1}{e^\varepsilon-1}.$$
Owing to the use of randomised response, this value in $[0,m]$ is converted into a single bit long report. The authors also provide a method to perform memoisation using one-bit estimation to protect rapidly updated data from longitudinal attacks.

\paragraph{LinkedIn Audience Engagement API} Rogers et al\cite{rogers2020linkedins} (2020) introduced a system to provide user-level privacy guarantees via differential privacy while being able to provide audience engagement insights to enable marketing analytics and related applications. In particular, the authors describe a number of DP algorithms (for cases where the data domain is reasonably sized and known, and where the data domain is unknown or very large in size) that help the LinkedIn Audience Engagement API to carry out privacy-preserving data analysis. They also introduce a privacy budget management system that tracks an analyst's privacy budget even over multiple data centres.
\subsection{By Uber}
Johnson et al\cite{Johnson2018TowardsPD} (2018), introduced key innovations to enable the practical use of differential privacy. Some of their most prominent contributions include the introduction of \emph{elastic sensitivity} which is a novel and convenient method to approximate and upper bound the local sensitivity, and can be used to obtain parameters to employ any local sensitivity based DP mechanism. Building on top of that, they propose FLEX, an end-to-end differential privacy solution for real-world SQL queries that uses elastic sensitivity.
\section{Bibliometric Analysis}
With a profusion of research in differential privacy and its various application being published in recent years, we shall provide some very brief bibliometric insights into the same to inform about trends in research and future directions.

For this, we shall be using the arXiv dataset, given that most significant works on differential privacy are available on arXiv. Starting from 2006, a total of 1653 papers have been published on arXiv with the term 'differential privacy' and case variations thereof in their abstracts, with a total of 1020 authors having published at least 2 papers on DP, and a total of 479 authors having published at least 3 papers. 
\begin{table}[!h]
\centering
\begin{tabular}{|l|c|l|l|c|}
\cline{1-2} \cline{4-5}
Statistic/Percentile & \multicolumn{1}{l|}{Value} &  & Statistic/Percentile                 & \multicolumn{1}{l|}{Value} \\ \cline{1-2} \cline{4-5} 
Mean                 & 1.834890                   &  & Mean           & 1.316706                                       \\ \cline{1-2} \cline{4-5} 
Standard Deviation   & 2.358568                   &  & Standard Deviation       & 0.788847                                       \\ \cline{1-2} \cline{4-5} 
Minimum              & 1                          &  & Minimum             & 1                                       \\ \cline{1-2} \cline{4-5} 
25\%                 & 1                          &  & 25\%         & 1                                       \\ \cline{1-2} \cline{4-5} 
50\%                 & 1                          &  & 50\%              & 1                                       \\ \cline{1-2} \cline{4-5} 
75\%                 & 2                          &  & 75\%     & 1                                       \\ \cline{1-2} \cline{4-5} 
Maximum              & 42                         &  & Maximum               & 9                                       \\ \cline{1-2} \cline{4-5} 
\end{tabular}
\caption{Authorship statistics by number of papers published on Differential Privacy (on the left) and that on Differential Privacy \emph{and} Machine Learning (on the right) (since 2005)}
\label{tab:BibStats}
\end{table}
The left subtable of table \ref{tab:BibStats} provides some simple statistics on authorship of these papers.

Now focusing on the literature published since 2012, it is observed that a total of 1609 papers were published with differential privacy being mentioned in their abstracts on arXiv, with a total of 998 authors having contributed to at least 2 papers, and with a total of 469 authors having contributed to at least 3 papers since 2012. This indicates that the bulk of research on differential privacy has occurred in the last decade.

Out of those 1609 works published since 2012, 418 papers feature the terms "machine learning", "gradient descent", "empirical risk", and "deep learning" (and case variations thereof) in their abstracts.

The right subtable of table \ref{tab:BibStats} provides statistics on authorship of papers per author on differential privacy mentioning topics related to machine learning in their abstracts
. Note that while the papers in this subset of the data mention machine learning, deep learning, and/or ERM, they might not deal with DPML, so the number of papers that actually deal with the applications of differential privacy to machine learning might be lesser than the number reported.

\begin{figure}[!h]
    \centering
    \includegraphics{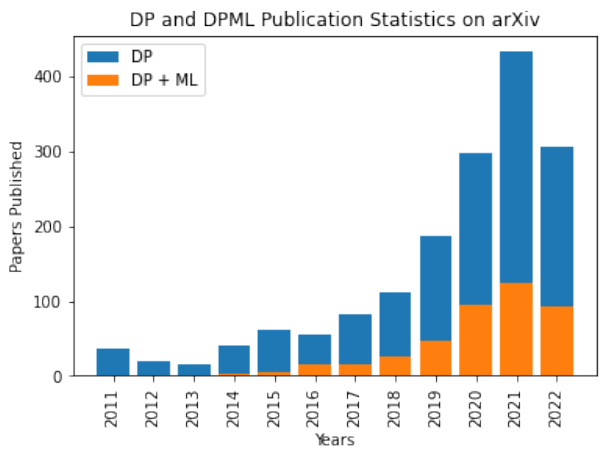}
    \caption{Bar Graph showing the number of DP and DPML publications in each year; the 2022 statistics are as of June, 2022.}
    \label{fig:dpvdpml_arxiv}
\end{figure}

Figure \ref{fig:dpvdpml_arxiv} depicts the number of papers mentioning DPML and those mentioning differential privacy as a whole since the years since 2011. It can be seen that the number of publications on differential privacy and machine learning with differential privacy have seen a consistent and significant increase in recent years, and that DPML has grown to account for a significant proportion of DP publications in the last 3-4 years.

 In addition, works on differentially private machine learning have become a staple of top AI conferences in recent years. Some data on this is available on \url{https://differentialprivacy.org} and the respective conference websites. For instance, NeurIPS 2020 featured 31 works on differential privacy, and NeurIPS 2021 featured 48 works on the same. ICML 2021 and ICML 2020 featured 21 and 22 works on differential privacy respectively. COLT 2020 featured 9 papers dealing with differential privacy.
\section{Remarks and Conclusion}
This survey seeks to be a reflection of the massive strides made in recent years in the field of differential privacy, and the various applications of the same. It brings the focus back to differential privacy and technicalities of the same; in particular, some prominent variants of differential privacy and differentially privacy techniques, accounting techniques and algorithms, and novel developments in terms of these were discussed. In addition, its real world applications from a differential privacy-first lens to fields like machine learning, deep learning, federated/distributed learning and DP-ERM were explored. We also discuss a few implementations of differential privacy in industry and for important tasks like census data privatisation.

This merely discusses a prominent subset of the profusion of research that has been done in differential privacy and its applications, and many of the techniques mentioned here have their own challenges in terms of feasibility of practical implementation, the privacy-utility tradeoff, the amount of data required to get high utility with high privacy, and improving on these remains the subject of much study. The goal of this survey is to augment existing survey literature on different facets/applications of differential privacy, and to show how differential privacy has become, and rightly so, the de-facto standard of privacy with wide ranging applications and implications.
\bibliographystyle{splncs04}
\bibliography{acmdp}

\appendix
\end{document}